# Atomistic deformation behavior of single and twin crystalline Cu nanopillars with preexisting dislocations

Won-Seok Ko [a,*], Alexander Stukowski [b], Raheleh Hadian [c], Ali Nematollahi [c], Jong Bae Jeon [d], Won Seok Choi [e], Gerhard Dehm [c], Jörg Neugebauer [c], Christoph Kirchlechner [c,f], and Blazej Grabowski [g]

[a] School of Materials Science and Engineering, University of Ulsan, 44610 Ulsan, Republic of Korea
[b] Department of Materials Science, Technical University of Darmstadt, 64289 Darmstadt, Germany
[c] Max-Planck-Institut für Eisenforschung GmbH, Max-Planck-Str. 1, 40237 Düsseldorf, Germany
[d] Advanced Surface Coating and Processing R&D Group, Korea Institute for Industrial Technology, 46938 Busan, Republic of Korea
[e] Department of Materials Science and Engineering, Korea Advanced Institute of Science and Technology, 34141 Daejeon, Republic of Korea
[f] Institute for Applied Materials, Karlsruhe Institute of Technology, Karlsruhe, Germany
[g] Institute of Materials Science, University of Stuttgart, Pfaffenwaldring 55, 70569 Stuttgart, Germany

**Abstract**

Molecular dynamics simulations are performed to investigate the impact of a coherent $\Sigma 3$ (111) twin boundary on the plastic deformation behavior of Cu nanopillars. Our work reveals that the mechanical response of pillars with and without the twin boundary is decisively driven by the characteristics of initial dislocation sources. In the condition of comparably large pillar size and abundant initial mobile dislocations, overall yield and flow stresses are controlled by the longest, available mobile dislocation. An inverse correlation of the yield and flow stresses with the length of the longest dislocation is established and compared to experimental data. The experimentally reported subtle differences in yield and flow stresses between pillars with and without the twin boundary are likely related to the maximum lengths of the mobile dislocations. In the condition of comparably small pillar size, for which a reduction of mobile dislocations during heat treatment and mechanical loading occurs, the mechanical response of pillars with and without the twin boundary can be clearly distinguished. Dislocation starvation during deformation is more pronounced in pillars without the twin boundary than in pillars with the twin boundary because the twin boundary acts as a pinning surface for the dislocation network.

*Corresponding author: Won-Seok Ko
wonsko@ulsan.ac.kr
Tel: +82 52 7128068

# 1. Introduction

Twin boundaries (TB) in fcc metals are known to improve both strength and ductility, although these features are usually considered mutually exclusive in common materials [1]. However, even today the simplest case of dislocation transmission through a Σ3 coherent TB is still not fully understood. For dislocation-TB interaction, two cases of dislocation slip transfer can be discerned: (i) Easy slip transfer of a screw dislocation, where the Burgers vector in the twin lies in the TB plane and likewise on a well-suited slip plane in the parent grain, and (ii) slip transfer with high breakthrough stresses, where the Burgers vector of the dislocation in the twin does not correspond to a low indexed lattice vector in the parent grain.

In the past decade, several theoretical studies have addressed the possible scenarios [2-22]. Jin *et al.* [2] investigated easy slip transfer of screw dislocations based on molecular dynamics (MD) simulations. In their work, a twin bi-crystal of either Al, Cu, or Ni material was relaxed under constant applied shear strain. Besides the transfer mechanism (i) outlined above, they [2] also observed dissociation of two partials moving in opposite directions along the twin boundary plane. Whether easy slip transfer or absorption occurs depends on a material specific shear stress threshold (e.g., 465 … 510 MPa for Cu [2]). The implicit limitations imposed by the used interatomic potentials as well as finite size effects were addressed by Chassagne *et al.* [9]. They [9] also confirmed a critical stress required to transmit dislocations and suppress absorption of the two partials in the twin boundary. Very recently, Dupraz *et al.* [22] performed improved MD simulations on twin bi-crystals of Al, Cu and Ni, overcoming the quasi 2-dimensional restriction of previous MD simulations [2, 9]. The calculated shear stress threshold of the slip transfer is significantly reduced (e.g., 260 MPa for Cu [22]) because the 3-dimensional geometry allows for the interaction between curved dislocation lines and the TB.

Experimental works have focused on slip transfer of individual dislocations through the TB in Cu [23-33]. Imrich *et al.* [25] showed that micron sized pillars containing a TB and loaded in a [101] direction exhibit an identical mechanical response as equally sized single crystals. The observed identical flow stresses could be in principle explained by the formation of dislocation pile-ups, amplifying the Peach-Köhler forces on the leading dislocation. However, *in situ* nanomechanical testing inside of a transmission electron microscope did not reveal any dislocation pile-ups [26, 27]. The same holds true for recent *in situ* Laue microdiffraction experiments by Malyar *et al.* [30] on TB containing micro pillars loaded in several directions between [101] and [112], where the effect of pile-ups was shown to be negligible. In Ref. [30], it was proposed that the required shear stress for slip transfer through a TB in Cu is independent of the loading direction and can be as low as 17 MPa, which is at least one order of magnitude smaller than the predictions by previous MD simulations [2, 22].

Despite the progress, a comprehensive understanding of the dislocation-TB interaction, combining in particular simulation and experimental results, is not yet available. A fundamental discrepancy is observed in the stress state where dislocation-TB interaction occurs. The origin can be traced back to a difference in the boundary conditions of the simulations and experiments. In usual experiments on pillar compression, multi-axial stress states may influence the dislocation-TB interaction, but previous MD simulations [2, 9] considered only a pure shear type deformation for the dislocation movement in periodic simulation cells along a designated slip plane. Moreover, these simulations did not capture the effect of transmission of multiple dislocations located on different slip planes.

To reduce the discrepancies between simulations and experiments a further MD simulation study [16] had been initiated focusing on similar geometries and loading conditions as in experiments. Specifically, Jeon and Dehm [16] performed compression tests of Cu pillars with the TB oriented analogous to the experimental compression tests [25-27, 30, 32, 33]. They [16] found that a dislocation network including sessile dislocations can form at the TB resulting in a strong orientation dependence. Interestingly, their work reports a clear difference between pillars loaded in [101] and [112] directions with only the former direction showing a pronounced formation of a dislocation network at the TB. Such a dislocation network with sessile dislocations is a potential barrier to further dislocation motion, and thus it can be expected that the TB loaded in the [101] direction shows hardening as loading progresses. However, experimental results are contradictory. The works by Imrich *et al.* [25, 26] compared flow stresses of Cu pillars with and without the TB loaded in the [101] direction, reporting only a marginal difference between them. The work by Malyar *et al.* [30] reported a similar trend as Imrich *et al.* [25, 26] also for [112], [123], [134], and [235] loading directions and no orientation dependence. Moreover, a later work by Malyar *et al.* [33] on 129 micropillars reported that Cu pillars with a TB deform on average only at slightly higher stress values with respect to the pillars without the TB loaded in the [347] direction.

These contradicting results between simulations and experiments indicate that there is still a critical difference in the boundary conditions. A potentially influential difference concerns the initial defect structure of the pillars before mechanical loading. In the usual experimental pillars, there is a possibility that preexisting dislocation sources are included from the initial macroscopic sample. Moreover, because experimental pillars are mostly prepared by the focused ion beam (FIB) technique, ion-induced defects such as vacancies and dislocations can be introduced in a region near the surface of the pillars [27]. This particular condition was not accounted for by the previous MD simulations [16] which conducted the compression on initially perfect and dislocation-free pillars. It is likely that preexisting dislocations influence the dislocation-twin boundary interaction. For example, a TB in the pillar might be locally faceted or kinked and defects introduced during the FIB milling can be expected to exist.

The present study focuses on MD simulations of dislocation-containing pillars to provide results that are more closely comparable to the experimental observations. This aim is achieved by adopting two different preparation techniques introduced previously by Sansoz [34] in deformation simulations of single-crystalline Cu nanopillars and by Zepeda-Ruiz *et al.* [35] in deformation simulations of bulk Ta. The work of Ref. [34] revealed that by introducing vacancies in the starting configuration and by conducting a high-temperature heat treatment, initial dislocations can be intentionally introduced into the pillar. It was shown [34] that the deformation of the pillar is significantly affected by the preexisting dislocations. Such a procedure is analogous to the experimental preparation of nanopillars at least for the surface near region, where a large number of defects is introduced by FIB preparation followed by partial healing due to heat treatment [26, 27]. Because this preparation technique inevitably overestimates the proportion of the FIB affected region due to the size limitation of MD simulations, we employ additionally the preparation technique developed in the work of Ref. [35], which introduces initial dislocations through vacancy-like prismatic loops placed on the possible slip planes. This technique readily provides a way to control the number of dislocation sources. However, the disadvantage is that dislocations can easily escape to the free surfaces, again due to the size limitation of MD simulations.

In the present study, we investigate the deformation behavior of Cu nanopillars with and without a TB utilizing both of the complementary preparation techniques of Ref. [34] and Ref. [35] to introduce preexisting dislocations, thus enabling MD compression tests that resemble experimental conditions. On this basis, a comprehensive discussion on the impact of the TB on the deformation of the pillars becomes possible. Moreover, we utilize the dislocation details obtained from a dislocation extraction algorithm (DXA) [36] to establish a relationship between the dislocation structures, in particular the maximum dislocation lengths and density, and the mechanical behavior of the pillars.

## 2. Methodology

Deformation simulations were performed for Cu pillars with a square-shaped cross-section as investigated in the experiments of Ref. [30]. Pillars with different lateral dimensions (11, 14, 21, 28, 35, and 42 nm) were prepared using the same aspect ratio of 1.414 (height/width). The lateral dimensions were treated as free surfaces, whereas periodic boundary conditions were applied along the loading axis to minimize the influence of the aspect ratio on the predicted deformation behavior. The resulting cell dimensions and numbers of atoms for each pillar are summarized in Table 1. To verify that the influence of the aspect ratio is small, the periodic length along the loading direction was doubled in a selected cell. Corresponding simulation results based on a large number of statistically independent runs (Supplementary Fig. S1) show that doubling the aspect ratio and thereby the height of the pillar has indeed a small effect on the average yield and flow stresses. The serration amplitudes during the plastic deformation decrease for the larger aspect ratio which can be explained by the larger number of *statistically independent* defects, i.e., not constrained by the periodic-boundary conditions, that stabilize the activation of new dislocations required for the plastic flow. This observation is consistent with the general results to be presented in the following sections. Nevertheless, it should be mentioned that, due to the periodic boundary conditions, the simulated Cu cell corresponds to an ideal, infinite wire thereby differing from realistic experimental pillars, which are often affected by misalignment and lateral constraints due to the friction between the pillar and indenter.

In order to investigate the effect of the TB on the deformation behavior, twin crystal pillars (TCPs) were generated as shown in Fig. 1. Following the previous experimental setup of Ref. [30], a coherent Σ3 TB was introduced at the center of the bi-crystal cell, running from edge to edge and parallel to the loading direction, and five crystallographic loading directions were investigated: [112], [235], [123], [134], and [101]. Corresponding single crystalline pillars (SCPs) were prepared with loading orientations identical to their bi-crystalline counter parts.

MD simulations were performed using the LAMMPS code [37] with an embedded-atom method (EAM) interatomic potential for pure Cu developed by Mishin et al. [38]. This potential had been developed based on experimental and density-functional-theory (DFT) data with the aim to accurately reproduce important properties closely related to the deformation of materials such as the stacking fault energy (44 mJ/m$^2$, cf., the experimental value of 45 mJ/m$^2$ [39]) and unstable stacking fault energy (158 mJ/m$^2$, cf., the DFT values of 158 mJ/m$^2$ [40] and 181 mJ/m$^2$ [41]). To elucidate a possible influence of the choice of the interatomic potential on the obtained simulation results, we have performed detailed benchmark simulations with another EAM potential, specifically the one developed by Sheng et al. [42], which exhibits a higher stacking fault energy (53 mJ/m$^2$) and unstable staking fault energy (190 mJ/m$^2$). The general physical properties predicted by both EAM potentials are summarized in the Supplementary Table S1. As for the specific deformation behavior, we focused on the evolution of the overall stress-strain response of differently sized [101] pillars, as presented in Supplementary Fig. S11 (Sheng potential) which should be compared with Fig. 8 below (Mishin potential). Comparison of the results for the two interatomic potentials reveals a very similar deformation behavior, thus indicating that the conclusions to be drawn in the following sections are independent of the specific choice of the interatomic potential.

All MD simulations in the present study were performed with a time step of 5 fs, which was previously shown [34] to be small enough to adequately capture the deformation behavior of Cu pillars using the chosen EAM potential [38]. The Nosé-Hoover thermostat and barostat [43, 44] were used for controlling temperature and pressure, respectively. Initially, the generated pillars were subjected to an energy minimization process using the conjugate gradient method to obtain equilibrium configurations.

The procedures introduced in Ref. [34] (referred to as "vacancy annealing method" in the following sections) and Ref. [35] (referred to as "vacancy loop method" in the following sections) were adopted to achieve an initial dislocation structure inside of the nanopillars before any straining. In the case of the vacancy annealing method, we randomly distributed certain amounts of vacancies (2, 5, and 10 %) in each SCP and TCP. Subsequent annealing at designated temperatures and times was then carried out based on an isobaric-isothermal (*NPT*) ensemble at zero-pressure as shown in Supplementary Fig. S2. This procedure is in close analogy to the annealing of FIB-caused knock-on damage as proposed in Ref. [45] and experimentally performed in Ref. [26]. In the case of the vacancy loop method, we randomly distributed certain numbers (24, 48, and 96) of vacancy loops with different diameters [(5, 10, 15, 20, 25, 30, 35, and 40) × *a* (0.3615 nm)] on each slip plane of each SCP and TCP. The position of the loops was randomly selected in each of the single crystals, and if a loop overlapped with the TB or the free surface it was clipped on one side. During a relaxation run at 300 K and zero-pressure for 100 ps, the distributed vacancy loops relaxed into dislocation lines forming a network. Details of the preparation process of pillars with an initial dislocation structure are presented in the Supplementary material [Sec. S-1 and Figs. S4–S5].

Compression tests of the pillars were conducted under a canonical (*NVT*) ensemble at 300 K. A strain-controlled uniaxial compressive loading was applied with an engineering strain rate of $1 \times 10^8$ s$^{-1}$ and a maximum strain of 15 % by controlling the cell dimension corresponding to the loading direction of the pillar.

The strain rate of the present simulations is in the conventional range of usual MD simulations ($10^7 - 10^9$ s$^{-1}$) [46], but is orders of magnitude higher than that of usual experiments on nanopillars ($\approx 10^{-3}$ s$^{-1}$ [25, 26]). To analyze the effect of the strain rate, benchmark simulations were performed for a simulation cell with a dimension of 28 nm using various, computationally accessible strain rates ($5 \times 10^6 - 1 \times 10^9$ s$^{-1}$). The resultant stress-strain response (Supplementary Fig. S3) indicates that the overall stress level is well converged for strain rates below $1 \times 10^8$ s$^{-1}$. Considering this convergence behavior and balancing it against the computational requirements, a strain rate of $1 \times 10^8$ s$^{-1}$ was selected and used for all simulations which will be presented in the following sections.

To visualize the evolution of the microstructure during the compressive loading, local atomic arrangements were identified using the Ackland–Jones analysis [47] as implemented in the LAMMPS code and the OVITO program [48]. Information on dislocation line lengths, density and Burgers vectors was extracted from the MD trajectories using the DXA dislocation analysis function of OVITO [49]. In our analysis individual dislocations were distinguished on the basis of their different Burgers vectors. The resultant atomic configurations and dislocation structures were visualized using OVITO.

## 3. Results
### 3.1. Deformation behavior of pristine nanopillars

To have a reference for investigating the impact of the initial dislocation structure on the compression, it is helpful to first examine the deformation behavior of pristine, i.e., dislocation free, nanopillars. Figure 2 shows the stress-strain response of differently sized pristine SCPs and TCPs for the loading orientations of [101] and [112] at 300 K. The results for the loading orientations of [235], [123], and [134] are shown in Supplementary Fig. S6. For all these dislocation-free pillars, we find three characteristic features that can also be expected in general for experimental nanowhiskers without dislocations. First, very high yield stresses (mostly ≈5 GPa and ≈10 GPa for the [101] loading) are observed for both SCPs and TCPs. Second, if yield stresses of differently sized pillars are compared, they are varying only stochastically and the well-known sample size effect (i.e., "smaller is stronger") is not present. Finally, a very serrated-like behavior in the stress-strain curves is observed.

These distinctive characteristics of pristine pillars are known and related to the deficiency of available dislocation sources inside of the pillars. Since the simulations discussed in this section were started without any initial dislocations, the initiation of the plastic deformation must be accompanied by the nucleation of new dislocations. The much higher yield stresses in the simulations compared to experiments can be explained by the fact that the stress required for the nucleation of a new dislocation should be much higher than that required for activating existing dislocations.

An attractive interaction exists between the dislocation line and the free surface which is caused by a surface image stress. This attraction implies that the first dislocation nucleation site should be at the corner of two free surfaces as it minimizes the dislocation length at the nucleation event, hence reducing the nucleation barrier. This is indeed the case in all our calculations of the pristine pillars (Supplementary Movies S1–S5). In particular, for the employed square-shaped pillars, the controlling factor for the first dislocation nucleation is independent of the size of the pillars, since the contact angle between two adjacent free surfaces is always the same. Interestingly, the [112], [235], [123], and [134] orientations show similar yield stresses of 5 − 6 GPa as indicated in Figs. 2(b, e) and Supplementary Fig. S6. The small stochastic variation for these orientations is likely caused by thermal vibrations. The [101] direction is special in that it requires a yield stress about twice as high (10 GPa) as shown in Figs. 2(a, d).

This feature can be explained by differences in the effective shear stress of each loading for the nucleation of partial dislocations. According to the previous study on the deformation of Au nanowires by Lee *et al*. [50], the dominant deformation mechanism is decisively controlled by the orientation-dependent resolved shear stress on the *leading* and *trailing* partial dislocations rather than that on the perfect dislocation. As shown in the Supplementary Movies (S1–S5), the nucleation of dislocations in the pristine pillars starts with a pair of a leading and trailing partial dislocation. Table 2 lists Schmid factors for the perfect <011> slip and for the partial <112> slip decomposed into values for the leading and trailing partial dislocation in each loading direction. The Schmid factors for the perfect <011> slip are similar for the different loading directions and thus cannot account for the extra high yield stress of the [101] loading. Instead, the Schmid factor for the leading partial dislocation of the [101] loading indicates a significantly lower value (0.236) than for the other loading orientations (0.314 – 0.337). We performed additional tensile loading simulation and confirmed that the yield stress is indeed decisively driven by the Schmid factor for the leading partial dislocation (see Fig. 3).

After yielding, the nucleated dislocation travels along the corresponding slip plane indicating a sudden drop in the stress value. The dislocation moves at comparably low stresses, annihilating as soon as it meets the free surface. Then, the nucleation of another dislocation is necessary for further plastic deformation which requires comparable stresses to the nucleation of the first dislocation. These repetitive processes of nucleation and disappearance of dislocations explain the serrated-like behavior in the stress-strain curve. Figure 2 shows that large pillars have generally weaker serration amplitudes compared to smaller sized ones. This can be explained by pronounced dislocation-dislocation interactions. In the larger sized pillars, coexistence of multiple dislocations is more likely because the dislocations need more time to escape from the pillar as the size increases.

### 3.2. Deformation behavior of nanopillars with preexisting dislocations

In the previous section, the deformation behavior of dislocation-free nanopillars was examined. However, in most micron and submicron sized metallic pillars many dislocations are already present before loading. In this section, results of MD simulations based on pillars with initial dislocations introduced by the preparation methods of Ref. [34] and Ref. [35] (Supplementary Figs. S4–S5) are presented. The same orientations and sizes of pillars as in the previous section (Sec. 3.1) are analyzed.

Figures 4(a) and (b) show a histogram of the dislocation length distribution in the [101] pillars prepared by the vacancy annealing method. It can be seen that the size distribution of the stair-rod type dislocations is confined to smaller dislocation lengths (up to about 5 nm), whereas the distribution of the Shockley partial type dislocations has a significant tail towards larger dislocation lengths. Frequently, Shockley partials of about 10 nm are found, however, some long partials can reach values of more than 17 nm. Figures 4(c) and (d) show a histogram of the dislocation length distribution in the [101] pillars prepared by the vacancy loop method. Compared to the results from the vacancy annealing method, longer Shockley partial and stair-rod dislocations can be obtained with a significantly smaller fraction of stair-rod dislocations. For the vacancy annealing method, the presence of the TB in the TCPs as well as the pillar orientation do not significantly affect the distribution of dislocations after annealing. Qualitatively, the similar distribution of dislocations in SCPs and TCPs after heat treatment is exemplified by the two representative snapshots in the insets of Figs. 4(a) and (b). In fact, there is even a good quantitative agreement in the dislocation length distribution as shown by the histograms in Figs. 4(a) (SCP) and (b) (TCP). The total dislocation density is, within the available statistics, independent of the presence of the TB and also of the loading orientation. Instead it is affected by the initial vacancy concentration, annealing time, and annealing temperature (Supplementary Fig. S9).

Figure 5 show the stress-strain response of the differently sized SCPs and TCPs prepared by the vacancy annealing method for the loading orientations of [101] and [112] at 300 K. The results for the loading orientations of [235], [123], and [134] are shown in Supplementary Fig. S7, and corresponding snapshots are shown in Supplementary Movies S6−S10. These results are significantly different from the deformation behavior of the pristine pillars considered in the previous section. First, the yield stresses of the larger sized pillars (35 and 42 nm) are significantly reduced (down to 1 − 2 GPa) and are now closer to the results of comparable experiments (0.4 – 0.95 GPa [26, 27]). There is still a difference in the overall magnitude, but we will show in Sec. 4 that the remaining discrepancy can be explained by considering the size difference between the simulation cells (15 − 42 nm) and experimental samples (400 – 650 nm [26, 27]). Second, the well-known "smaller is stronger" effect is now observed in the stress-strain responses of pillars with different sizes. Finally, the serration amplitude is greatly reduced in the stress-strain response of the larger sized pillars (35 and 42 nm), and it is similar to experiments on pillars with initial dislocations [25, 30].

Figure 6 shows the stress-strain response of the 42 nm sized SCPs and TCPs prepared by the vacancy loop method for the loading orientations of [101] and [112] at 300 K. The results for the loading orientations of [235], [123], and [134] are shown in Supplementary Fig. S8, and corresponding snapshots are shown in Supplementary Movies S11−S15. Similar to the deformation behavior of pillars prepared by the vacancy annealing method, the yield stresses are significantly reduced compared to the deformation behavior of the pristine pillars. However, the serration amplitude, especially for the SCPs, is generally higher than for the deformation behavior of pillars prepared by the vacancy annealing method for the same sized pillars of 42 nm. The serration amplitude is reduced when the size of initial vacancy loops increases.

## 4. Analysis of the impact of the TB on the yield stress and correlation with the maximum dislocation length

The yield stress corresponds to the point in the stress-strain curve at which the first dislocation becomes active. It is therefore governed by the distribution of initial dislocations. The pristine pillars are in this respect an extreme case with zero initial dislocations and, as already mentioned, at yielding new dislocations need to be generated with the edges of the free surfaces acting as the preferential nucleation sites. This finding implies that the perfectly flat TB does not provide any lower-energy nucleation sites. An interesting question is, whether the TB influences dislocation nucleation in the region where it hits the edge of the pillar (cf. Fig. 1). In all our calculations of the [112], [235], [123], and [134] orientations, we find that the dislocations nucleate at the edges without the TB (Supplementary Movies S2–S5). This means that the TB increases the nucleation stress, i.e., an edge+TB geometry has higher nucleation stress than a pure edge, the latter being in the range of $5 - 6$ GPa [Figs. 2(b, e) and Supplementary Fig. S6]. As explained in Sec. 3.1, the nucleation of dislocations starts with a pair of a leading and trailing partial dislocation, and the yield stress is determined by the Schmid factor of the leading partial dislocation. Since a dislocation nucleating at the edge with the TB would need to constrict at the TB and locally form a perfect screw dislocation, such a nucleation process would in fact require an elevated nucleation stress level comparable to that of a perfect dislocation. For the [101] orientation we observe that dislocation nucleation is possible for both a pure edge and an edge with TB (Supplementary Movie S1). Thus, here the edge+TB geometry has a similar nucleation stress to the pure edge [yield stresses twice as high at 10 GPa; Figs. 2(a, d)], and stresses required for the nucleation of the partial and perfect dislocations in this orientation are comparable. The specific behavior of the [101] orientation appears to be related to the particularly low Schmid factor of the leading partial dislocation (Table 2).

We turn to the impact of the TB on the yield stress of pillars with initial dislocations, a situation close to experimental conditions [26, 27]. We concentrate in particular on the stress-strain curves most closely resembling experimental data, i.e., the ones for the larger pillars in the vacancy annealing method (Fig. 5) and the ones with larger initial vacancy loops for the vacancy loop method (Fig. 6). Comparing the yield stresses of the SCPs [Figs. 5(a, b) and 6(a, b)] with the corresponding TCPs [Figs. 5(d, e) and 6(d, e)], we observe similar values in the range of $1 - 2$ GPa. Upon a closer look, the TCPs appear to have slightly higher yield stresses (in the range of a few hundred MPa) than the SCPs. Due to this small difference and the statistical variation of the yield stress which depends on the exact initial conditions, we have performed multiple independent stress-strain simulations considering both preparation methods for the [112], [123], and [101] loading orientations employing different initial conditions for the pillars, in order to produce statistically significant results. In the simulations with pillars prepared by the vacancy annealing method, we have used different vacancy concentrations (2, 5, and 10 %), annealing temperatures (500, 700, 900, 1100, and 1300 K), and annealing times (550, 750, 1000, 1500, and 2500 ps) for the preparation procedure. These different conditions resulted in a wide spectrum of total dislocation densities and maximum dislocation lengths after the heat treatment (Supplementary Figs. S9 and S10). In the simulations with pillars prepared by the vacancy loop method, we have used different numbers (24, 48, and 96) and widths [(5, 10, 15, 20, 25, 30, 35, and 40) × $a$ (0.3615 nm)] of initial vacancy loops for the preparation procedure. The resulting averaged yield stresses for the TCPs, SCPs, and their difference including the variance are shown in Table 3. The larger set of calculations confirms that the TCPs have indeed slightly higher yield stresses than the SCPs regardless of the preparation method, with the difference being most pronounced for the [101] orientation.

In order to reveal the physical basis of this difference, we have analyzed the yield stress data assuming the weakest-link model (aka source truncation model) for which one expects that the longest dislocation is activated first [51]. This model was suggested to explain the observed size effect on the strengthening of materials in small dimensions with a focus on the stochastic nature of the length of dislocation sources. Within this model, the stress required for the activation of existing dislocations scales inversely with the dislocation source size and we therefore plot the yield stresses versus the maximum source sizes in Fig. 7. We have confirmed that the type of the dislocation with the maximum length in each pillar is always the mobile Shockley partial. This is because the Shockley partial dislocations very actively move and merge to larger dislocations during the heat treatment. An inverse correlation is visible for all the three [112], [123], and [101] loadings and for both, SCPs and TCPs. We have investigated other related factors such as the dislocation density and the average dislocation length, but found no clear correlation with the yield stress. This finding indicates that the longest mobile dislocation is a main factor determining the yield stress of the pillars irrespective of the presence of the TB.

Some deviation from the inverse correlation can be observed for the largest maximum dislocation lengths (around 20 − 30 nm) that result from the vacancy loop preparation method. In particular, we can see a strengthening effect (i.e., increase in the yield stresses) correlated with the loading orientation ([101] > [123] > [112]). Considering the initial dislocation structures that result from the two preparation methods the strength increase seems contradictory. As explained in the Supplementary material [Sec. S-1 and Figs. S4–S5], the pillars prepared by the vacancy annealing method exhibit a significantly higher dislocation density and larger amounts of stacking fault tetrahedra compared to those prepared by the vacancy loop method. One can expect higher yield stresses for higher dislocation densities due to forest hardening [52, 53]. Moreover, the role of stacking fault tetrahedra on preventing dislocation motion was revealed by previous MD [54-58] and discrete dislocation dynamics [59] studies. The here found hardening behavior for the largest maximum dislocation lengths obtained with the vacancy-loop method thus suggests that there is an additional factor that determines the yield stress, in addition to the maximum dislocation length.

A possible additional factor is the detailed distribution of the initial Shockley partial dislocations. The pillars prepared by the vacancy annealing method (Supplementary Movies S6–S10) generally exhibit a significantly higher stacking fault area between the leading and trailing partial dislocation (colored atoms visualized by the Ackland–Jones analysis [47]). Thus, the yielding of the corresponding pillars is decisively driven by the activation of *trailing* partial dislocations as well as *leading* partial dislocations. In contrast, as the mobile dislocations in the pillars prepared by the vacancy loop method are mostly full dislocations (pairs of closely coupled leading and trailing partial dislocations), the yielding of the respective pillars is governed by the activation of long *leading* partial dislocations (Supplementary Movies S11–S15). The calculated Schmid factors under compressive loading (Table 2) indicate that the effective shear stress for trailing partial dislocations is greater than that for leading partial dislocations. The increased yield stress of pillars prepared by the vacancy loop method can be then explained by the generally smaller Schmid factors of leading partial dislocations. The orientation dependency of the hardening behavior ([101] > [123] > [112]) can be likewise explained by the difference in Schmid factors of leading and trailing partial dislocations ([101] > [123] > [112]) as listed in Table 2.

The virtue of the representation employed in Fig. 7 is that we can quantitatively compare with experiment. Unfortunately, experimental data linking the dislocation source size length directly to the flow strength are scarce. For dislocations in the nanometer regime data does – to the best of our knowledge – not exist. However, Imrich and co-workers imaged several larger dislocation sources with a size of the order of 100 nm during operation *in situ* during pillar compression inside a TEM [26]. This allows for a direct correlation of the dislocation source activation stress to the initial dislocation source size. The corresponding experimental data fall well onto the inverse correlation as derived from the simulation data based on the weakest-link model (Fig. 7). This agreement indicates that the maximum dislocation length is a main factor controlling the experimental yield stress.

Based on this understanding, we have investigated whether we can also relate the small difference in the SCP and TCP yield stresses discussed above (Table 3) to the maximum dislocation length. One may expect that the TB leads to shorter maximum dislocation lengths and thus to higher yield stresses than in the corresponding SCPs if the maximum dislocation length scales with the grain size (i.e., a half of the pillar dimension). Although this scenario applies to pillars with penetrable high angle grain boundaries [60], a recent experiment by Malyar *et al*. [33] reported that the maximum dislocation length does not scale with the grain size in the pillars with the TB. Instead, they proposed a presence of double-hump dislocations in the TCPs to explain the subtle differences in the SCP and TCP yield stresses. According to this hypothesis, the maximum lengths of dislocations in SCPs and that in TCPs extending through the TB are comparable. Because the dislocations extending through the TB have to constrict at the TB to form a perfect screw dislocation, they should exhibit a double-hump shape with an increased curvature that results in an increased activation stress.

We have confirmed that the expected double-hump dislocations are indeed present during the deformation of TCPs (Supplementary Movies S1–S15). Further, in agreement with the experimental prediction and as discussed in Supplementary material (Sec. S-1), the present MD simulations cannot detect any clear difference between the initial dislocation length distribution in SCPs and TCPs. In fact, the maximum dislocation length varies statistically rather strongly in all pillars, such that sometimes the SCPs have larger maximum dislocation lengths and sometimes the TCPs (Supplementary Fig. S10). This variation is reflected in Fig. 7 by the fact that the data points do not lie on a sharp line, i.e., there is no perfect correlation. The corresponding spread on the stress axis is larger than the difference in the yield stresses between SCPs and TCPs, and this is why correlating this difference with the maximum dislocation length is not possible based on the present data.

We can thus conclude that the maximum dislocation length is a main factor in determining the yield stress,

but other factors are important for capturing the details. We expect these factors to be (i) the specific form of the initial dislocations, i.e., whether full dislocations (a set of adjacent leading and trailing partials) or separated leading and trailing partial dislocations are dominant, (ii) the exact orientation of the longest dislocation with respect to the loading axis, (iii) the intrinsic resistance of the TB to dislocation transmission, (iv) the increased curvature of double-hump dislocations extending through the TB, and (v) the additional resistance due to the agglomeration of dislocations close to the TB. The last point is supported by the fact that the largest difference between SCP and TCP yield stresses is found for the [101] orientation (Table 3) which also shows the most pronounced tendency for the agglomeration of dislocations at the TB and, in addition, by the prominent role of dislocation networks close to the TB in understanding the flow stress as discussed in the following.

## 5. Analysis of the impact of the TB on the plastic deformation behavior
### 5.1. Impact of the TB on the flow stress of the larger pillars

We now move on to analyze the impact of the TB on the flow stress, i.e., the stress required to continue plastic deformation after yielding. For the pristine pillars we observe strong serrations of the flow stresses (Fig. 2 and Supplementary Fig. S6) even for the largest pillar size of 42 nm. The serrations have a similar magnitude for the SCPs and TCPs, and it is difficult to detect any influence of the TB. An exception is the [101] TCP [Fig. 2(d)], for which the flow stresses seem on average smaller than in the [101] SCP [Fig. 2(a)] and where the flow stresses of the 35 and 42 nm pillars fluctuate around a value of 2 GPa beyond a strain of 0.08. This special feature of the [101] orientation is related to the formation of a dislocation network at the TB [16] as will be investigated in more detail in Sec. 5.4.

Regarding the flow stress curves of pillars with initial dislocations, we concentrate on the pillars prepared by the vacancy annealing method (Fig. 5) and, among these, first on the larger sized (42 nm) pillars which show very stable flow stress curves with respect to deformation, closely resembling experimental data. The pillars prepared by the vacancy loop method exhibit higher flow stresses with strong serrations especially for the case of the SCPs, thereby preventing a direct comparison between SCPs and TCPs in terms of the average flow stress. To quantify the impact of the TB on the flow stress, we have increased the statistics by performing 30 independent stress-strain curves using the same preparation parameters (vacancy concentrations, annealing temperatures and times) as in Sec. 4 for the yield stress calculations. We have calculated average flow stresses for all stress-strain curves in a strain interval from 0.05 to 0.15. The resulting flow stresses were further averaged over the different conditions for each compression orientation and are shown in Table 3. The larger set of calculations reveals that the TCP flow stresses are higher by 0.16 − 0.26 GPa than for the SCPs.

We will relate most of the observed flow stress features to the maximum dislocation length during deformation (Secs. 5.2 and 5.3) and also to the formation of dislocation networks (Sec. 5.4).

**5.2. Correlation of the flow stress with the length of the longest mobile dislocation**

For the yield stress we found an inverse correlation with the maximum dislocation length available (Fig. 7). To see whether the flow stress also shows a similar correlation, we have analyzed in detail the dislocation length distributions during the deformation behavior of the nanopillars with preexisting dislocations prepared by the vacancy annealing method (corresponding snapshots are shown in Supplementary Movies S16−S21). Figure 8 shows the corresponding results for a subset of the investigated nanopillars. The maximum dislocation lengths (blue dots and black squares) were obtained for strain intervals of 0.5 % and are superimposed in Fig. 8 on the respective stress-strain curves.

A comparison of the maximum dislocation lengths with the flow stresses suggests that there is a correlation between these quantities. Moreover, one can see that smaller maximum dislocation lengths result in a more serrated behavior of the flow stress. For example, for the smallest SCP of 11 nm [black symbols and line in Fig. 8(a)] we observe on average the shortest maximum dislocation lengths (about 3 nm) and correspondingly the strongest serrations in the flow stress and also the highest peaks (about 7 GPa). For the same sized TCP, longer dislocation sources are found (roughly 10 nm) resulting in reduced serrations and flow stress peaks (about 3 GPa). When the pillar size increases, larger maximum dislocation sources become available. The corresponding flow stresses show less serrations and smaller peak values. For the two largest pillars the maximum dislocation lengths fluctuate from about 20 to 40 nm and the flow stresses are rather stable showing only small fluctuations at around 2 GPa.

In order to better quantify the inverse correlation between the flow stress and the maximum dislocation length, we have averaged the flow stresses of each pillar in the strain interval from 0.05 to 0.15 and plotted them versus the corresponding average maximum dislocation length. As shown in Fig. 9, the inverse correlation between the flow stress and the average maximum dislocation length is clearly visible, similarly as found above for the yield stress. In fact, the source truncation model was originally suggested by analyzing the size effect of the *yield strength* originating from the stochastic nature of the dislocation source lengths in samples of finite size [51]. Our results suggest that the activation of the longest dislocation source governs the overall deformation of pillars not only at the yield point but also further during the deformation process irrespective of the presence of a TB.

**5.3. Size dependent deformation behavior of pillars with initial dislocations: Dislocation starvation**

For the pillars with initial dislocations prepared by the vacancy annealing method (Fig. 5 and Supplementary Fig. S7), the qualitative behavior of the stress-strain curves depends on the pillar size. The smaller pillars (11, 14, and 21 nm) exhibit higher flow stresses with strong serrations. For these smaller pillars, the [101] orientation is special in that the SCPs have larger flow stresses than the respective TCPs. This result appears to deviate from the usual expectation that a TB should lead to a higher flow stress or at least to a similar one if it does not act as a barrier. In fact, we have already verified in Sec. 5.1 with a statistically significant estimate that the TB indeed increases the deformation resistance for the larger pillars. As discussed in the following, the here observed, counter-intuitive deformation characteristic of the smaller pillars (11, 14, and 21 nm) for the [101] orientation can be interpreted as a result of dislocation starvation [61] in combination with the stabilizing role of the TB (Sec. 5.4).

Owing to the large surface-to-volume ratio of the smaller pillars, mobile dislocations can comparably easily escape from the pillar during the heat treatment and the subsequent deformation process. Due to this starvation of mobile dislocations, the flow stress increases and exhibits strong serrations because new dislocations need to be nucleated at higher stress levels. As revealed by Figs. 8(a)–(c), the smaller sized [101] SCPs (11, 14, and 21 nm) show higher flow stresses and serration amplitudes than the corresponding TCPs, because dislocation starvation is more pronounced in the SCPs compared to the TCPs in which the TB acts as a stabilizer of the dislocation network.

For the larger pillars [35 and 42 nm; Figs. 8(e) and (f)], dislocation starvation is suppressed during the deformation, in both SCPs and TCPs. For these sizes, the overall flow stresses and serrations of the stress-strain curves of the SCPs and TCPs are comparable, because the long mobile dislocations are retained in the pillars during the heat treatment and subsequent deformation. The stabilization occurs here not only via the TB (for the TCPs) but also due to dislocation-dislocation interaction (for both TCPs and SCPs). As a consequence, the flow stresses are rather stable showing only small fluctuations and the mechanical response of these pillars is in the regime of the source truncation model, where the overall deformation is controlled by the truncated length of existing mobile dislocations.

In the medium sized pillar (28 nm) in the [101] orientation [black curve in Fig. 8(d)], an intermediate behavior with a special stress-strain curve is obtained for the SCP: A relatively stable flow stress over a certain amount of strain and then a sudden change to a serrated behavior. At the initial stage of the deformation, a weak serration amplitude is observed in the stress-strain curve meaning that the pillar deforms with sufficient dislocation sources. After a significant loading of $\varepsilon \approx 0.08$, the serration amplitude is drastically increased and we interpret this as a sudden starvation of available dislocations during the deformation. Our argument is supported by the evolution of the densities of dislocation components as a function of strain as shown in Fig. 10(a) (open symbols). As the deformation progresses, the overall dislocation density gradually decreases (a feature possibly related to "mechanical annealing" [62]), but at a strain of around 0.08 there is a drastic decrease especially in the Shockley partial dislocations. It seems that theses mobile dislocations are freed from the pinning points inside of the pillar and are able to escape to and out of the surfaces. Further deformation requires the activation or nucleation of dislocations at a higher stress level.

## 5.4. Formation of dislocation networks and the stabilizing role of the TB

The present MD simulations reveal that TCPs exhibit a different behavior with respect to dislocation starvation than SCPs. In addition to the above discussed results for the SCP density evolution, Fig. 10 also shows densities of dislocation components in the 28 and 42 nm sized [101] TCPs prepared by the vacancy annealing method. The density of Shockley partial dislocations in both TCPs does not change significantly but remains at a similar level as the density of stair-rod type dislocations. This result clearly indicates that sudden dislocation starvation is significantly suppressed in TCPs. This can be interpreted as a stabilizing effect of the TB on the dislocation network, providing secure dislocation sources throughout the deformation of the pillars. In fact, Jeon and Dehm [16] found previously that the [101] orientation readily forms a dislocation network near the TB during deformation. We have thus investigated the formation of dislocation networks in the present context of pillars with initial dislocations and the various loading orientations.

We focus on the dislocation structure of the 42 nm TCPs after compression ($\varepsilon$ = 0.15). The results for all investigated loading orientations are shown in Fig. 11 and we observe the importance of the initial cell preparation on the formation of the dislocation networks. Figures 11(a)-(e) show pristine TCPs corresponding to the initial conditions of the simulations in Ref. [16]. In consistency with Jeon and Dehm's finding [16] we observe that the pristine [101] TCP forms a dislocation network close to the TB [Fig. 11(a)], but not the pristine [112] TCP [Fig. 11(b)]. For the here additionally investigated [235], [123], and [134] loading orientations, we find no dislocation networks for the pristine pillars. Hence it is only the [101] orientation that is special, forming a dislocation network due to the activation of dislocations where easy cross-slip for passing the TB is prevented. This finding explains why the larger sized pristine [101] TCPs showed distinct flow stresses among all pristine pillars, which stabilized at a value of about 2 GPa after a larger strain (≈0.08) with small remaining fluctuations. The dislocation network provides a statistically significant number of dislocation pinning points and sources that can be activated rather easily at similar stresses, thus resulting in small serrations in the flow stress dependence.

The situation changes when we consider the pillars with initial dislocations. Figures 11(f)–(j) show the dislocation structures formed after compressing these pillars to the same strain as the pristine pillars ($\varepsilon$ = 0.15). All loading orientations show now a tendency for the formation of dislocation networks with a higher dislocation density at and near the TB. Nevertheless, the [101] orientation still stands out with the most pronounced dislocation network, in particular in the vicinity of the TB.

The formation of the dislocation network for the [101] orientation was discussed in Ref. [16]. For the other orientations, dislocation networks can be caused by imperfections of the TB originating from the initial dislocation preparation procedure. A recent MD study by Fang *et al*. [19] showed that kinks at the TB can promote dislocation absorption to the TB, which is likely contributing to the formation of dislocation networks. Moreover, dislocation-dislocation interactions—strongly enhanced due to the initial dislocation content—should be an important factor contributing to the formation of a dislocation network.

The presence of dislocation networks explains why larger sized TCPs with initial dislocations have small serrations in their flow stresses regardless of loading orientation. The argument is the same as stated above for the pristine [101] TCP: A dislocation network provides enough dislocation sources that can be activated at relatively small stresses, and a sufficient number of these sources is available preventing an exhaustion during continued deformation.

Dislocation networks do not only form in the TCPs. The larger sized SCPs with initial dislocations also show a clear tendency to form dislocation networks as shown in Figs. 11(p)–(t). The mechanism responsible for dislocation networks in SCPs is dislocation-dislocation interaction. Interestingly, this mechanism alone leads to dislocation networks that are less stable than dislocation networks in the presence of a TB, resulting in less stable stress-strain curves as discussed in the previous section (Sec. 5.3).

Smaller sized pillars, whether SCPs or TCPs, form much less pronounced dislocation networks as exemplified in Fig. 12. This is the reason why the corresponding stress strain curves show significant serrations. Although these smaller sized pillars show a smaller tendency for the formation of dislocation networks compared to the larger pillars, the TB still provides sufficient pinning points to stabilize longer dislocations, which are activated during the deformation.

## 6. Conclusions

Large-scale MD simulations have been performed to provide a dislocation-based understanding of the deformation of Cu nanopillars with and without a twin boundary. Our work clearly reveals that the role of the twin boundary on the overall mechanical response of nanopillars is critically dependent on the availability of initial dislocation sources. In the condition of comparably large pillar sizes and abundant initial mobile dislocations, overall yield and flow stresses are decisively driven by the longest, available mobile dislocation irrespective of the presence of the twin boundary. In the condition of comparably small pillars, the twin boundary contributes to the deformation by preventing the starvation of mobile dislocations. Our results can be summarized as follows:

(1) The deformation of dislocation free pristine nanopillars by MD simulations shows features that are expected and experimentally confirmed for dislocation free pillars: They show a yield stress close to the theoretical strength. New dislocations need to be generated at the edges of the square-shaped pillars, which are the preferential nucleation sites for dislocations in the pristine pillars.

(2) The applied preparation procedure of the pillars to introduce a preexisting dislocation structure before the deformation is crucial to achieve a stress-strain response with features as observed in experiment. The vacancy annealing method provides a way to obtain very stable flow stress curves resembling previous experimental data on pillars prepared by FIB milling. The vacancy loop method complements the vacancy annealing method, providing a way of preparing pillars with preexisting dislocations that can be controlled in size and density.

(3) We have revealed that the longest, available mobile dislocation is the key quantity to determine the stress-strain response of nanopillars, irrespective of the presence of the twin boundary, while other factors are important for capturing the details. The resulting yield and flow stresses show an inverse correlation with the mobile dislocations of maximum length.

(4) For the larger investigated pillar sizes (35, 42 nm) with abundant preexisting dislocations, the mechanical response of the pillars is in the regime of the source truncation model, i.e., the deformation is controlled by the maximum length of any mobile dislocation. The experimentally reported subtle effect of the twin boundary on the stress-strain response can be qualitatively explained by the marginal difference in the maximum length of the mobile dislocations between the single crystal pillars and the twin boundary pillars. However, a unique quantitative correlation could not be established.

(5) For the smaller investigated pillar sizes (11, 14, 21 nm) with preexisting dislocations, dislocation starvation is a governing mechanism influencing the overall deformation behavior. During the thermal annealing and mechanical loading processes, mobile dislocations are freed from their pinning points and can escape through the surface. The flow stress increases along with the starvation of comparably long mobile dislocations because new dislocations are required to be nucleated/activated at higher stresses.

(6) The present MD simulations reveal a stabilizing effect of the twin boundary on the dislocation network of smaller sized nanopillars. Due to this effect, for the smaller investigated pillar sizes where dislocation starvation is the governing mechanism, the overall flow stress is lower in the twin crystalline pillars than in the single crystalline pillars. In particular, this is because the dislocation networks, which continuously provide dislocation sources, are stabilized at the twin boundary. For larger pillar sizes the twin boundary acts as a stabilizer as well, but its effect is less visible because dislocation-dislocation interaction leads to an additional stabilizing mechanism.

**Acknowledgements**

We thank Nikolay Zotov for fruitful discussions. The funding by the European Research Council (ERC) under the EU's Horizon 2020 Research and Innovation Programme (Grant No. 639211) is gratefully acknowledged by BG, while GD thanks the ERC for funding of GB-CORRELATE (Grant No. 787446). This research was also supported by the National Research Foundation of Korea (NRF) funded by Ministry of Science and ICT (Grant No. NRF-2019M3D1A1079214, NRF-2019M3E6A1103984, and NRF-2019R1F1A1040393). This research was also supported by the National Supercomputing Center with supercomputing resources including technical support (Grant No. KSC-2018-CHA-0017).

Table 1  Cell dimensions and numbers of atoms in the nanopillar cells considered in the present MD simulations.

| Cell dimension (nm) | 10.6×10.6 ×15.0 | 14.1×14.1 ×20.0 | 21.2×21.2 ×30.0 | 28.3×28.3 ×40.0 | 35.4×35.4 ×50.0 | 42.4×42.4 ×60.0 |
|---|---|---|---|---|---|---|
| Number of atoms | $0.15×10^6$ | $0.34×10^6$ | $1.2×10^6$ | $2.7×10^6$ | $5.4×10^6$ | $9.2×10^6$ |

Table 2  Schmid factors of possible slip systems in the fcc structure under compressive loading. For each orientation, the leading and trailing partial dislocations are identified according to the previous work by Copley and Kear [63].

| Loading direction | Perfect dislocation | Partial dislocation | | |
|---|---|---|---|---|
| | | Leading partial * | Trailing partial * | Difference |
| [112] | 0.408 | 0.314 | 0.393 | 0.079 |
| [235] | 0.451 | 0.335 | 0.447 | 0.112 |
| [123] | 0.467 | 0.337 | 0.471 | 0.135 |
| [134] | 0.471 | 0.326 | 0.490 | 0.163 |
| [101] | 0.408 | 0.236 | 0.471 | 0.236 |

* Schmid factors of leading and trailing partial dislocations under tensile loading are reversed.

Table 3  Averaged yield and flow stresses including the standard error for a large set of independent stress-strain curves for the 42 nm nanopillar using different preparation methods and initial conditions (see Sec. 4 and Sec. 5.1 for details). The $\Delta_{\text{TCP-SCP}}$ values correspond to the average and the error over individual differences between TCP and SCP at the same initial conditions.

| Property | Method | | [101] | [112] | [123] |
|---|---|---|---|---|---|
| Yield stress (GPa) | Vacancy annealing | SCP | 2.12±0.20 | 1.71±0.11 | 1.69±0.11 |
| | | TCP | 2.28±0.19 | 1.76±0.10 | 1.76±0.10 |
| | | $\Delta_{\text{TCP-SCP}}$ | 0.16±0.05 | 0.05±0.02 | 0.07±0.02 |
| | Vacancy loop | SCP | 1.87±0.12 | 1.38±0.09 | 1.42±0.09 |
| | | TCP | 2.05±0.11 | 1.58±0.07 | 1.63±0.08 |
| | | $\Delta_{\text{TCP-SCP}}$ | 0.18±0.05 | 0.20±0.04 | 0.21±0.06 |
| Flow stress (GPa) | Vacancy annealing | SCP | 1.67±0.05 | 1.29±0.02 | 1.36±0.04 |
| | | TCP | 1.93±0.01 | 1.52±0.02 | 1.52±0.03 |
| | | $\Delta_{\text{TCP-SCP}}$ | 0.26±0.04 | 0.23±0.02 | 0.16±0.02 |

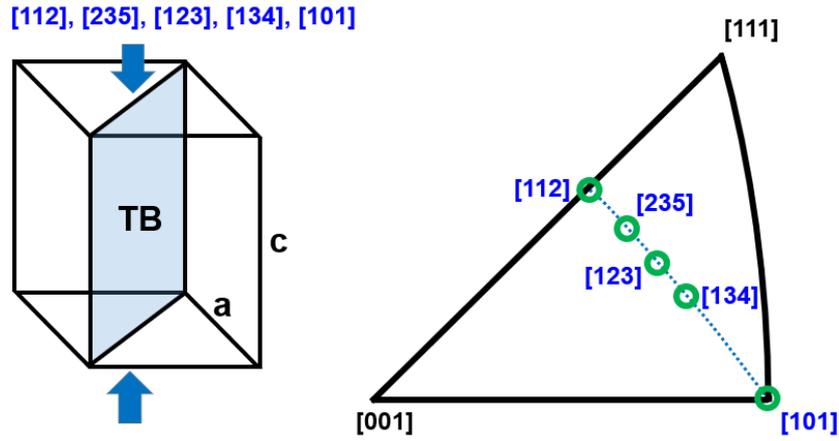

Fig. 1  Schematic illustration of the sample geometry and inverse pole figure showing compression directions for the MD simulations.

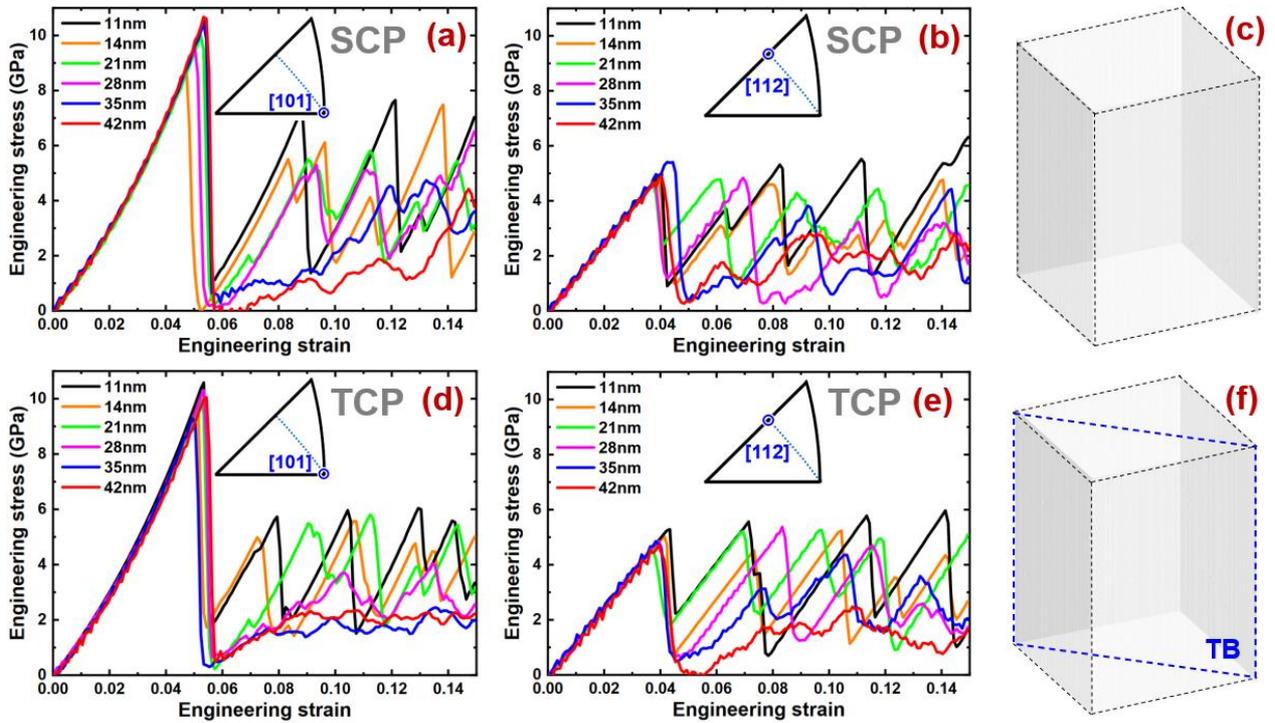

Fig. 2  Stress-strain response of initially dislocation-free (a, b) SCPs and (d, e) TCPs at 300 K. Results for various pillar sizes (11, 14, 21, 28, 35, and 42 nm) and different loading orientations of (a, d) [101] and (b, e) [112] are presented. (c) and (f) represent examples of initial configurations of the [112] SCP and TCP, respectively.

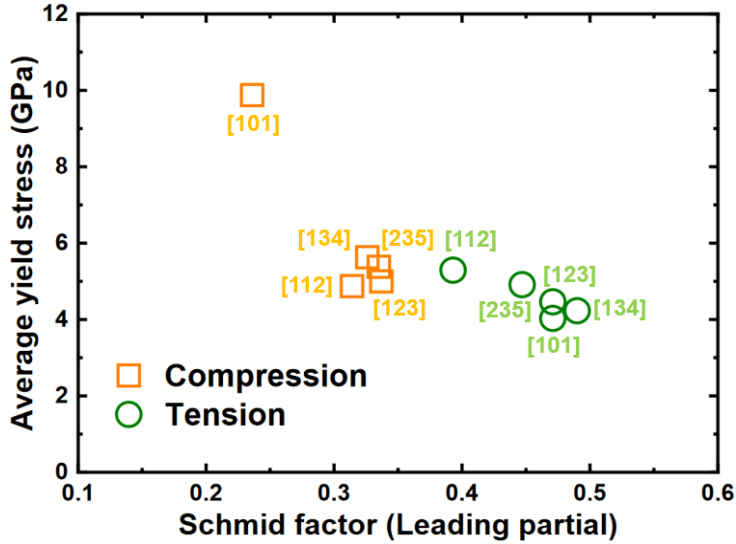

Fig. 3  Correlation between the Schmid factor of the leading partial dislocation and the average yield stress of pristine pillars obtained from compressive and tensile loading at 300 K. The yield stress values were calculated by averaging independent MD runs for differently sized (11, 14, 21, 28, 35, and 42 nm) SCPs and TCPs with the different loading orientations ([101], [112], [235], [123], and [134]).

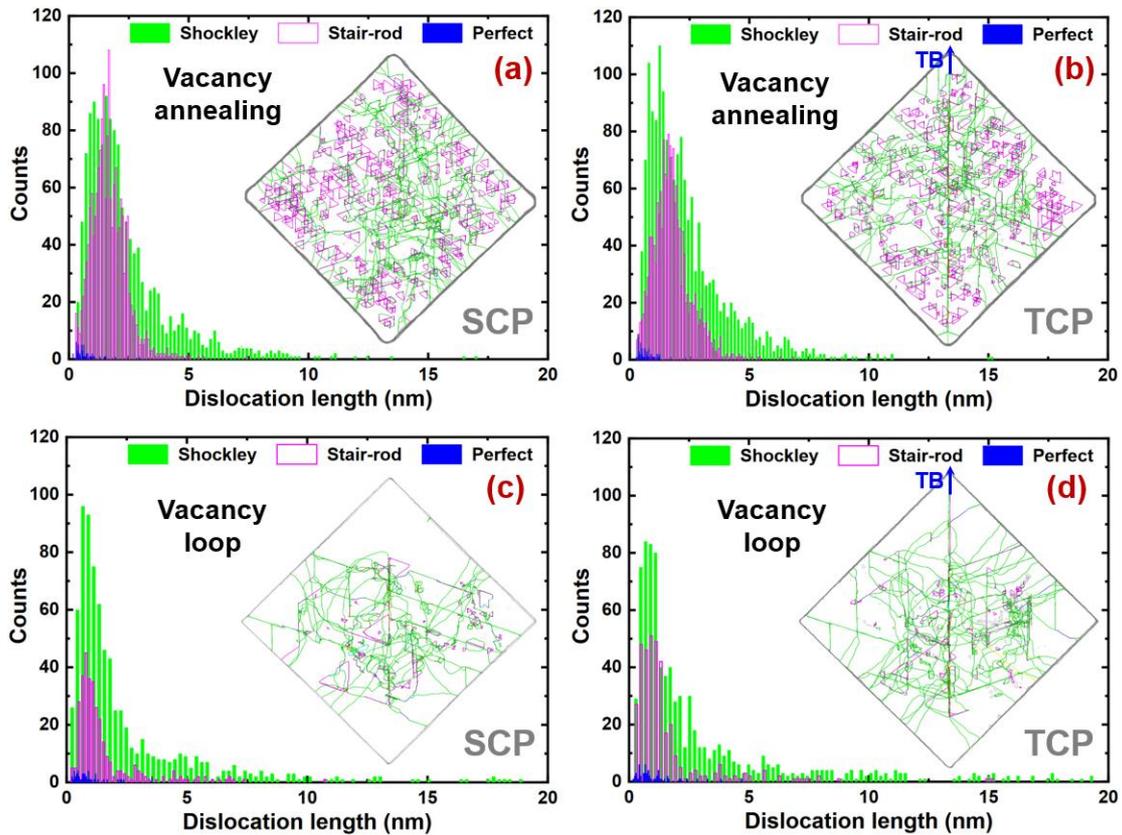

Fig. 4  Histograms of the dislocation length distribution for the 42 nm sized [101] SCPs and TCPs prepared by the two different methods. (a, b) Pillars with an initial vacancy concentration of 10 % were obtained after a heat treatment of 2500 ps at 1100 K. (c, d) Pillars with initially 24 vacancy loops with a width of $40 \times a$ (0.3615 nm) were relaxed at 300 K for 100 ps. Representative DXA snapshot are included, with green lines corresponding to Shockley partial dislocations, magenta lines to stair-rod dislocations, and blue lines to perfect dislocations.

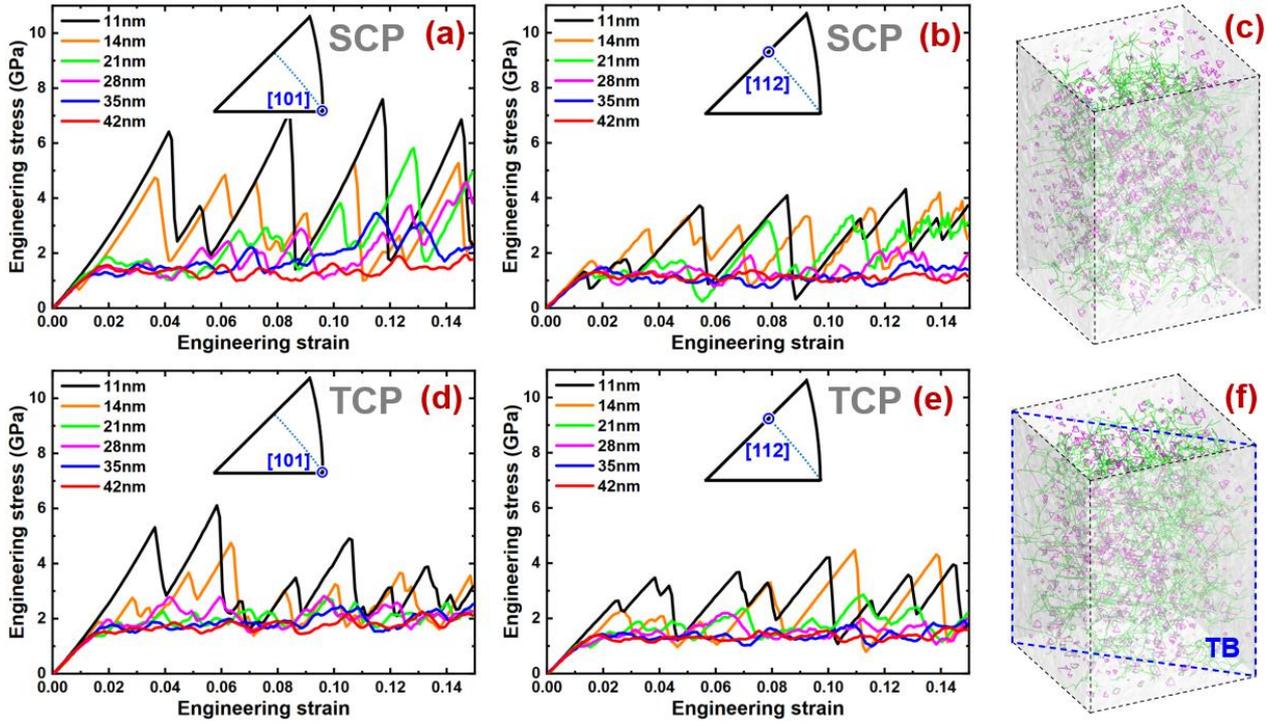

Fig. 5   Stress-strain response at 300 K of (a, b) SCPs and (d, e) TCPs prepared by a heat treatment of 750 ps at 1100 K with an initial vacancy concentration of 10 %. Results for various pillar sizes (11, 14, 21, 28, 35, and 42 nm) and different loading orientations of (a, d) [101] and (b, e) [112] are presented. (c) and (f) represent examples of initial configurations of the [112] SCP and TCP, respectively.

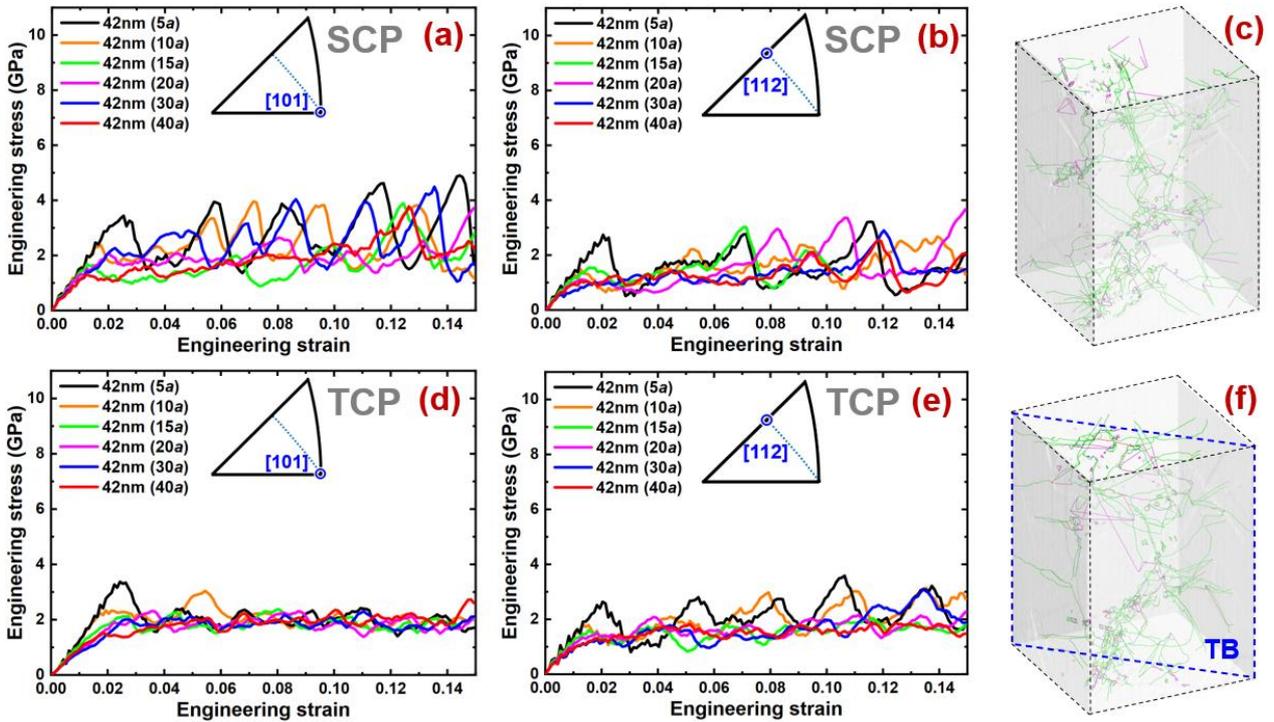

Fig. 6   Stress-strain response at 300 K of 42 nm sized (a, b) SCPs and (d, e) TCPs prepared by a relaxation of 100 ps at 300 K with initial 24 vacancy loops with different widths of (5, 10, 15, 20, 30, and 40) × $a$ (0.3615 nm). Results for different loading orientations of (a, d) [101] and (b, e) [112] are presented. (c) and (f) represent examples of initial configurations (40 × $a$) of the [112] SCP and TCP, respectively.

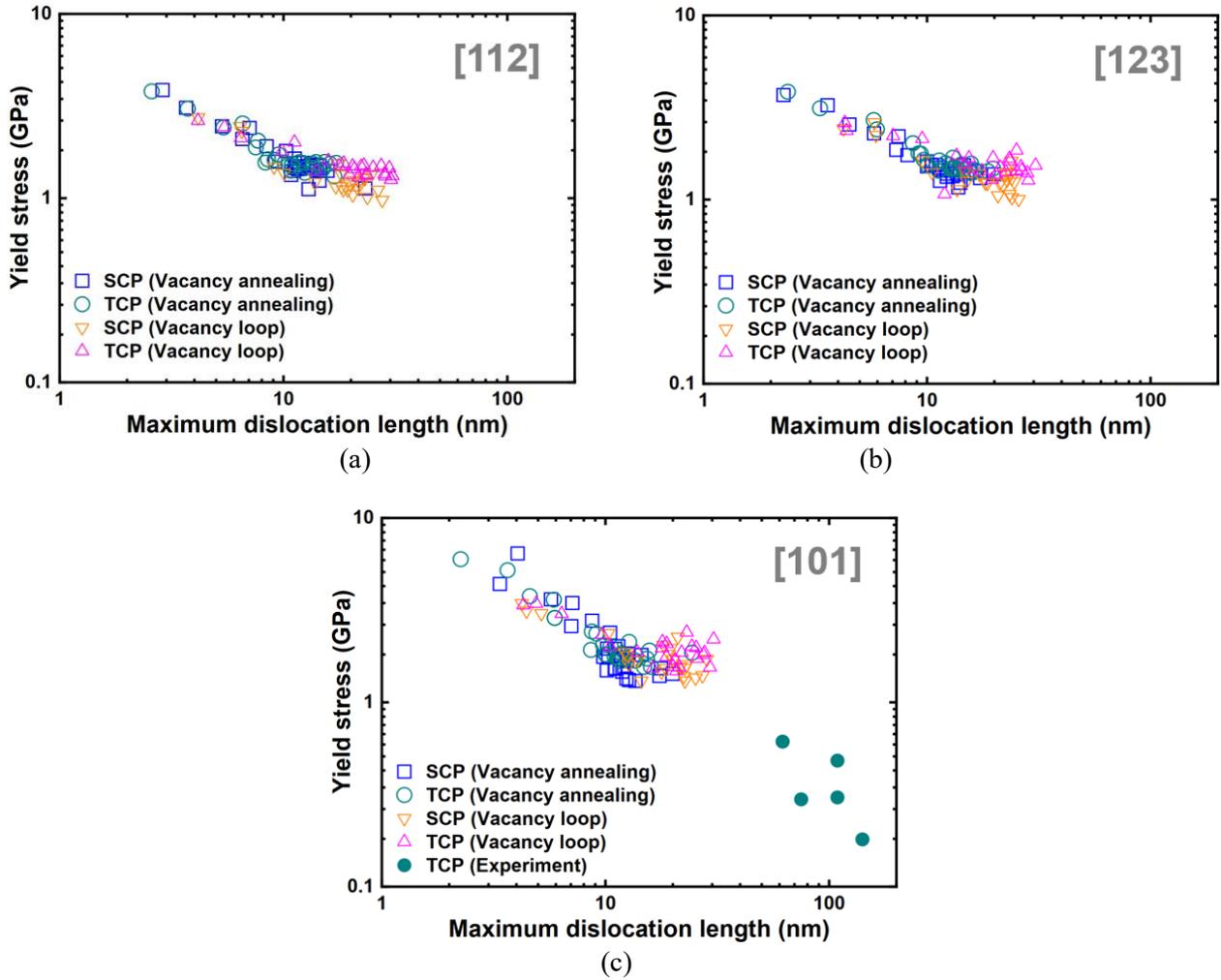

Fig. 7  Correlation between the maximum dislocation length and the yield stress obtained from the present MD simulations for the 42 nm sized nanopillars for the (a) [112], (b) [123], and (c) [101] loading directions, in comparison with experimental data (Ref. [26]). The various data points were obtained by compressing pillars prepared by the vacancy annealing and the vacancy loop method. The pillars prepared by the vacancy annealing method experienced various conditions of heat treatment (cf. Supplementary Fig. S2) and vacancy concentrations (2, 5, and 10 %). The pillars prepared by the vacancy loop method contained various numbers and widths of initial vacancy loops (Supplementary Table S2).

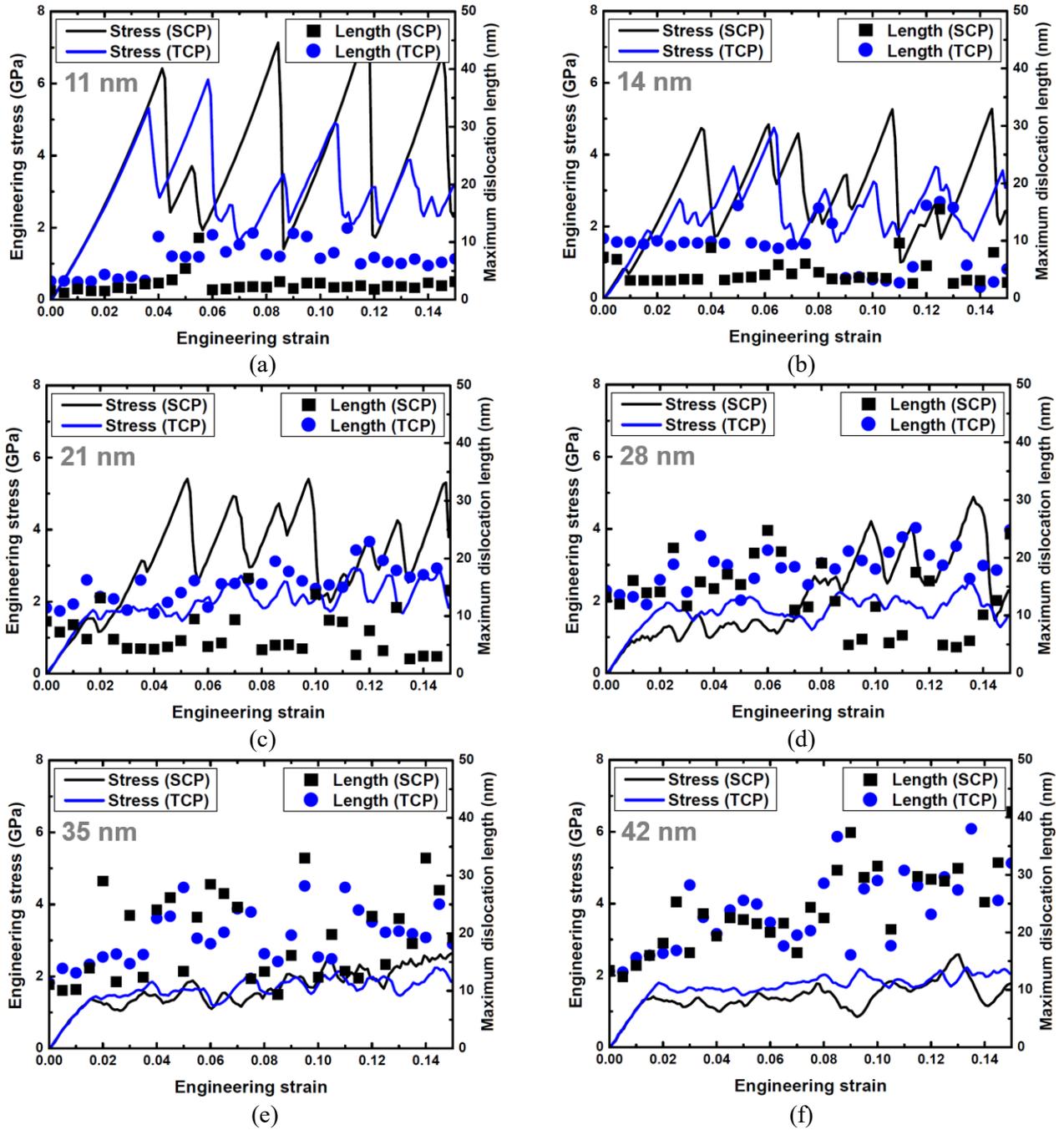

Fig. 8 Correlation between the flow stress and the maximum dislocation length for the (a) 11, (b) 14, (c) 21, (d) 28, (e) 35, and (f) 42 nm sized [101] SCPs and (b) TCPs during compressive loading at 300 K. Each pillar was prepared by the vacancy annealing method with an initial vacancy concentration of 10 % and heat treated for 750 ps at 1100 K before loading. Corresponding results based on the EAM potential by Sheng *et al.* [42] are presented in Supplementary Fig. S11.

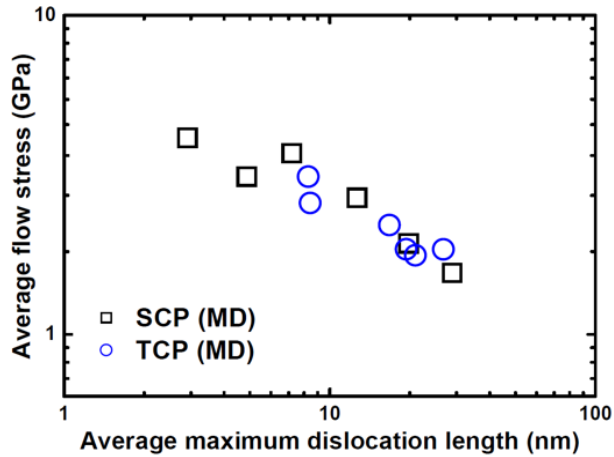

Fig. 9 Correlation between the average flow stress and the average maximum dislocation length of variously sized (11, 14, 21, 28, 32, and 42 nm) [101] SCPs and TCPs. The flow stresses and the maximum dislocation lengths for every 0.5 % strain intervals were averaged over a range of 5 – 15 % strain. Each point represents an average value for differently sized SCPs and TCPs, prepared by the vacancy annealing method with an initial vacancy concentration of 10 % and heat treated for 750 ps at 1100 K before loading.

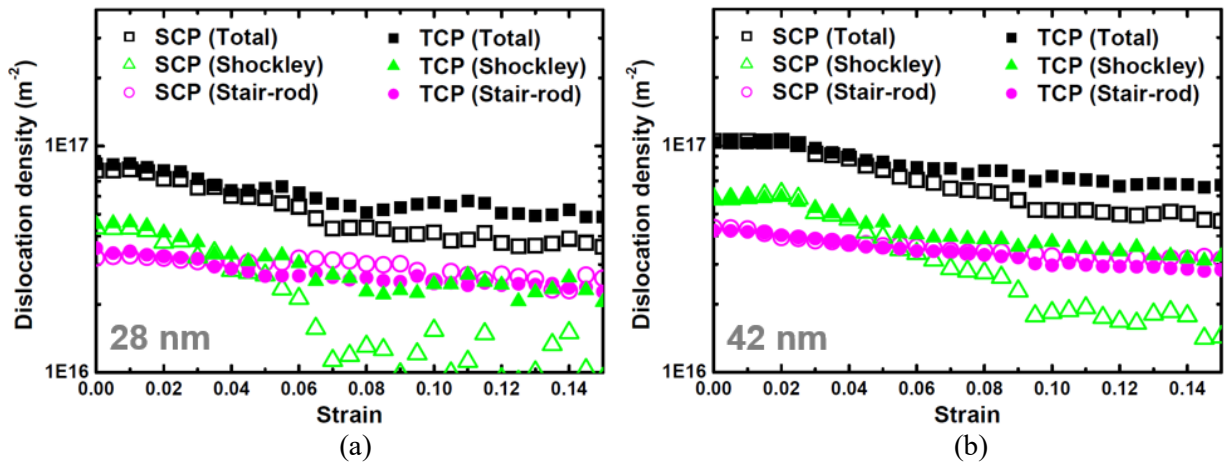

Fig. 10 Density of dislocation components obtained from the compressive loading of (a) 28 and (b) 42 nm sized [101] SCPs and TCPs at 300 K. Each pillar was prepared by the vacancy annealing method with an initial vacancy concentration of 10 % and heat treated for 750 ps at 1100 K before loading.

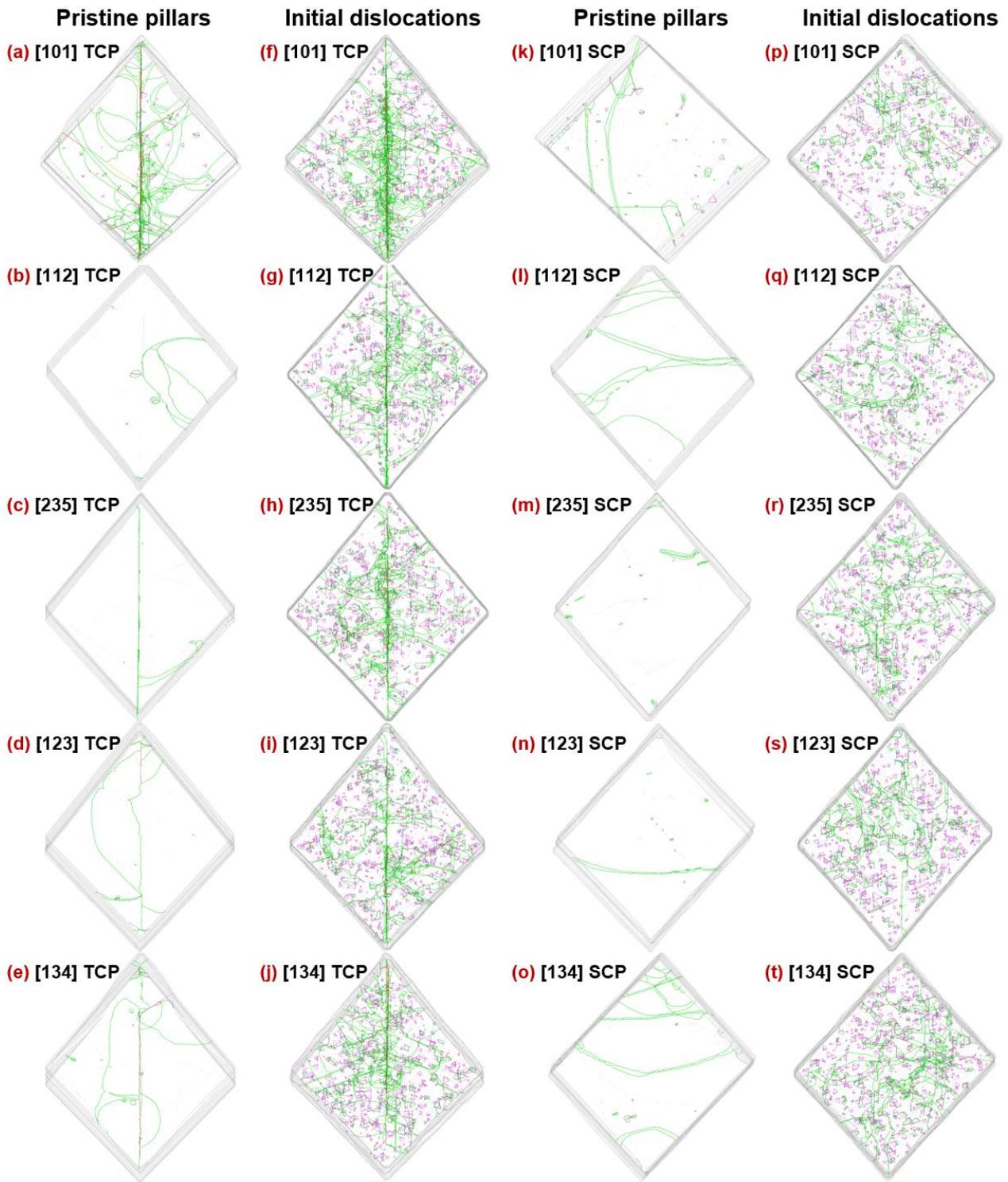

Fig. 11 Dislocation networks formed after compression ($\varepsilon$ = 0.15) of the 42 nm sized TCPs (a)–(e) without and (f)–(j) with initial dislocations and SCPs (k)–(o) without and (p)–(t) with initial dislocations prepared by the vacancy annealing method (10 % vacancies, annealing at 1100 K for 750 ps). Color schemes for dislocation types are as follows: green=Shockley partial dislocations, magenta=stair-rod dislocations, blue=perfect dislocations, and red=dislocations of other types.

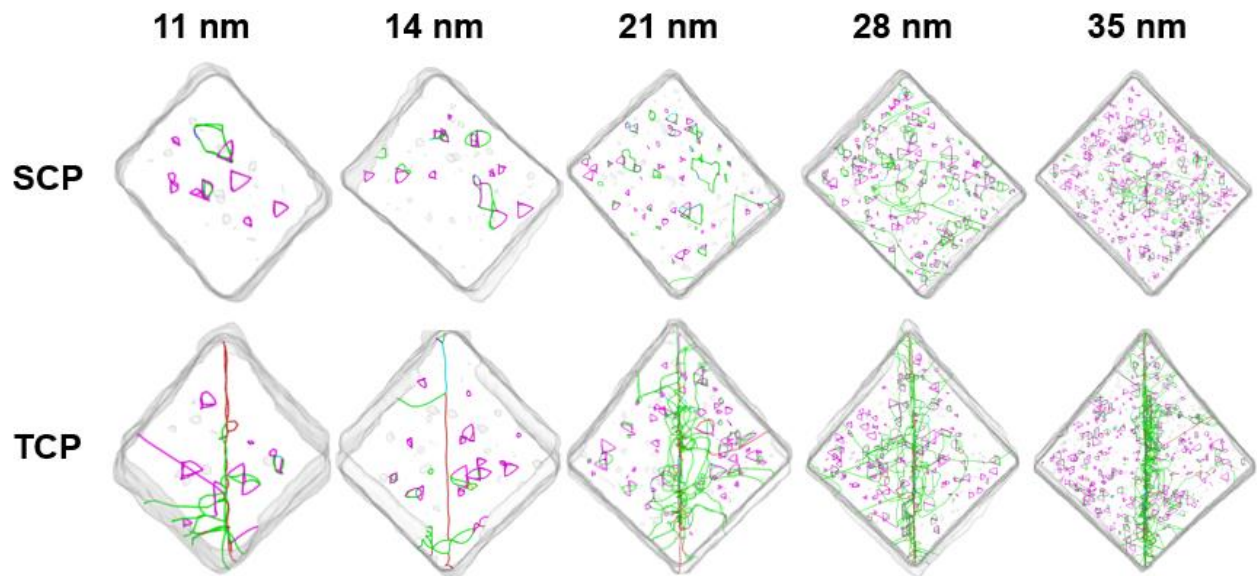

Fig. 12 Dislocation networks formed after compression ($\varepsilon$ = 0.15) of variously sized (11, 14, 21, 28, and 35 nm) [101] SCPs and TCPs with initial dislocations prepared by the vacancy annealing method (10 % vacancies, annealing at 1100 K for 750 ps). Color schemes for dislocation types are as follows: green=Shockley partial dislocations, magenta=stair-rod dislocations, blue=perfect dislocations, and red=dislocations of other types.

# Supplementary Material to "Atomistic deformation behavior of single and twin crystalline Cu nanopillars with preexisting dislocations"

Won-Seok Ko [a,*], Alexander Stukowski [b], Raheleh Hadian [c], Ali Nematollahi [c], Jong Bae Jeon [d], Won Seok Choi [e], Gerhard Dehm [c], Jörg Neugebauer [c], Christoph Kirchlechner [c,f], and Blazej Grabowski [g]

[a] School of Materials Science and Engineering, University of Ulsan, 44610 Ulsan, Republic of Korea

[b] Department of Materials Science, Technical University of Darmstadt, 64289 Darmstadt, Germany

[c] Max-Planck-Institut für Eisenforschung GmbH, Max-Planck-Str. 1, 40237 Düsseldorf, Germany

[d] Advanced Surface Coating and Processing R&D Group, Korea Institute for Industrial Technology, 46938 Busan, Republic of Korea

[e] Department of Materials Science and Engineering, Korea Advanced Institute of Science and Technology, 34141 Daejeon, Republic of Korea

[f] Institute for Applied Materials, Karlsruhe Institute of Technology, Karlsruhe, Germany

[g] Institute of Materials Science, University of Stuttgart, Pfaffenwaldring 55, 70569 Stuttgart, Germany

# S-1. Supplementary description: Details on the simulations

Figure S1 shows the result of benchmark simulations to examine the influence of the aspect ratio of pillars. Figure S2 shows the heat treatment process used for the vacancy annealing method. Figure S3 shows the result of benchmark simulations to examine the influence of the strain rate.

Fig. S4 shows the evolution of the dislocation structure in [101] pillars during the heat treatment process (cf. Fig. S2) of the vacancy annealing method visualized by OVITO. In the upper part [Fig. S4(a)-(d)] snapshots of the dislocation structure are presented for the example of the 42 nm sized [101] TCP with an initial vacancy concentration of 10 %, heat treated at 1100 K for up to 2500 ps. At the initial stage of the heat treatment [50 ps; Fig. S4(a)], the pillar contains numerous complex defect structures formed by the aggregation of vacancies. As the heat treatment progresses [300 and 2500 ps in Figs. S4(b) and (c)], initial defects gradually merge into longer dislocations. A detailed process is shown in Supplementary Movie S22 as an example. The resultant image of the pillar [2500 ps; Fig. S4(c)] shows a similar defect morphology as experimental TEM images of pillars prepared by FIB with subsequent heat treatment [1, 2]. In Figs. S4(e) and (f), the density of different dislocation types during the heat treatment is shown for an SCP and TCP, respectively, as obtained from the OVITO software. The density is defined as the length of all dislocations of a certain type divided by the volume of the unit cell. As the heat treatment progresses, the densities of all dislocation types gradually decrease resulting in a decrease of the total dislocation density. Mostly Shockley partial and stair-rod dislocations are present. The relative importance of these dislocations to other types (Frank, Hirth, and perfect) increases as the heat treatment progresses. We can see that the stair rod dislocations are associated with stacking fault tetrahedra as visualized in Fig. S4(d). Such tetrahedra are stable during the heat treatment suppressing further growth of the stair-rod type dislocations.

Figure S5 shows the evolution of the dislocation structure in the 42 nm sized [101] pillars prepared by the vacancy loop method. Results for the pillars with an initial number of 24 vacancy loops with widths of $15 \times a$ (0.3615 nm) and $40 \times a$ (0.3615 nm) are visualized as examples. Due to the relaxation of atomic positions, the "vacancy platelets" initially carved into the crystal transform into dislocation loops, which subsequently react with nearby loops to form a network. The density of the dislocations remaining after relaxation increases with increasing size and number of the initial vacancy loops, as summarized in the Table S2. Similarly to the dislocation structure obtained by the vacancy annealing method, most of dislocations are composed of Shockley partial and stair-rod types, but the amount of stair-rod dislocations and stacking fault tetrahedra is significantly decreased.

Figure S6 shows the stress-strain response of differently sized pristine pillars for the loading orientations of [235], [123], and [134] at 300 K. Figure S7 shows the stress-strain response of the differently sized pillars prepared by the vacancy annealing method for the loading orientations of [235], [123], and [134] at 300 K. Figure S8 shows the stress-strain response of the 42 nm sized pillars prepared by the vacancy loop method for the loading orientations of [235], [123], and [134] at 300 K. Figures S9 and S10 show the total dislocation densities and the maximum dislocation lengths of 42 nm sized pillars prepared by the vacancy annealing method, respectively.

The total dislocation density prepared by the vacancy annealing method is in the range of $2 \times 10^{16} - 4 \times 10^{17}$ m$^{-2}$ (Fig. S9), consistent with the previous MD study on Cu nanopillars [3], which also measured the dislocation density (after compression) using the DXA. However, the value is higher than the experimentally measured density in Cu nanopillars ($\approx 10^{14}$ m$^{-2}$) [4] and that of Ni nanopillars ($\approx 10^{15}$ m$^{-2}$) [5]. Lower dislocation densities, down to $\approx 7 \times 10^{15}$ m$^{-2}$, can be achieved with the vacancy loop preparation method. However, very small numbers and sizes of vacancy loops are required to obtain such low dislocation densities that are comparable to experimental values. For these low dislocation densities ($7 \times 10^{15}$ m$^{-2}$) the loading simulations do not reproduce the experimentally reported stress-strain response, instead they exhibit very serrated behavior with unusually large stress values in the stress-strain curves due to a deficiency of mobile dislocations as explained in Sec. 3.2.

The apparent conflict between simulations and experiment can be partially explained by considering differences in the estimation of the dislocation density. Whereas an accurate estimate is possible in the simulations due to the availability of all the relevant atomistic data, experiments are limited in measuring the dislocation density, which likely leads to an underestimation of the true density. For example, in experiments, leading and trailing partial dislocations are identified as full dislocations while the partial dislocations can be separated in the simulation data. Further, the experimental determination of dislocations inside of a pillar is usually estimated by TEM observations along the lateral direction of the compression. Therefore, small sized dislocations inside of a pillar, in particular inside of the FIB affected region, cannot be detected. In contrast, in atomistic simulations every dislocation, even the many short ones accumulated in highly defective crystal regions, can be identified and counted thanks to the DXA. However, it should be clarified that, beyond the experimental limits in the determination of the dislocation density, differences in the sizes of the pillars, in the boundary conditions along the loading axis, and in the heat treatment conditions (in particular the annealing time) between the MD simulations and the experiments likewise contribute to differences in the deformation behavior.

Figure S11 shows the evolution of the overall stress-strain response of differently sized [101] pillars performed by the different EAM potential (Sheng potential in Table S1) [7].

## S-2. Supplementary tables

Table S1  Calculated bulk and defect properties of pure Cu using the indicated EAM potentials [6, 7], in comparison with experimental and DFT data. Following quantities are listed: the cohesive energy $E_c$ (eV/atom), the lattice constant $a$ (Å), the bulk modulus $B$ and the elastic constants $C_{11}$, $C_{12}$ and $C_{44}$ (GPa), the structural energy differences $\Delta E$ (eV/atom), the vacancy formation energy $E_f^{\text{vac}}$ (eV), the vacancy migration energy $E_m^{\text{vac}}$ (eV), the surface energies $E_{\text{surf}}$ (mJ/m²) for the orientations indicated by the superscripts, the stable ($\gamma_{sf}$) and unstable ($\gamma_{us}$) stacking fault, and the twin fault ($\gamma_{tf}$) energies (mJ/m²).

| Property | Exp. | DFT | EAM [6] (Mishin) | EAM [7] (Sheng) |
|---|---|---|---|---|
| $E_c$ | 3.54 [a] | | 3.54 | 3.54 |
| $a$ | 3.615 [b] | | 3.615 | 3.615 |
| $B$ | 140 [c] | | 138.3 | 141 |
| $C_{11}$ | 176 [c] | | 169.9 | 175 |
| $C_{12}$ | 125 [c] | | 122.6 | 124 |
| $C_{44}$ | 82 [c] | | 76.2 | 79 |
| $\Delta E_{\text{fcc}\to\text{bcc}}$ | 0.04 [d] | | 0.046 | 0.04 |
| $\Delta E_{\text{fcc}\to\text{hcp}}$ | 0.006 [d] | | 0.008 | 0.01 |
| $E_f^{\text{vac}}$ | 1.03 − 1.30 [e] | | 1.272 | 0.99 |
| $E_m^{\text{vac}}$ | 0.65 [e] | | 0.689 | 0.74 |
| $E_{\text{surf}}^{(100)}$ | 1790 [f, i] | | 1345 | 1504 |
| $E_{\text{surf}}^{(110)}$ | 1790 [f, i] | | 1475 | 1607 |
| $E_{\text{surf}}^{(111)}$ | 1790 [f, i] | | 1239 | 1387 |
| $\gamma_{sf}$ | 45 [g] | 36 [j], 41 [k] | 44.4 | 53 |
| $\gamma_{us}$ | | 158 [j], 181 [k] | 158 | 190 |
| $\gamma_{tf}$ | 24 [h] | 18 [j], 20 [k] | 22.2 | 27 |

[a] Ref. [8].
[b] Ref. [9].
[c] Ref. [10].
[d] Ref. [11].
[e] Ref. [12].
[f] Ref. [13].
[g] Ref. [14]
[h] Ref. [15]
[i] The experimental value is for a polycrystalline solid.
[j] Ref. [16]
[k] Ref. [17].

Table S2  Density of dislocation components in the 42 nm sized [101] TCPs prepared by the vacancy loop method.

| Number of vacancy loops | Size of vacancy loop ($N \times 0.3615$ nm) | Initial dislocation density (m$^{-2}$) | | | |
|---|---|---|---|---|---|
| | | Total | Shockley partial | Stair-rod | Perfect |
| 24 | 5  | $6.76 \times 10^{15}$ | $2.68 \times 10^{15}$ | $3.80 \times 10^{15}$ | 0 |
|    | 10 | $8.16 \times 10^{15}$ | $5.77 \times 10^{15}$ | $1.93 \times 10^{15}$ | 0 |
|    | 15 | $1.44 \times 10^{16}$ | $9.75 \times 10^{15}$ | $3.82 \times 10^{15}$ | $2.06 \times 10^{13}$ |
|    | 20 | $1.71 \times 10^{16}$ | $1.32 \times 10^{16}$ | $2.39 \times 10^{15}$ | $1.27 \times 10^{14}$ |
|    | 25 | $1.79 \times 10^{16}$ | $1.31 \times 10^{16}$ | $3.36 \times 10^{15}$ | $1.08 \times 10^{14}$ |
|    | 30 | $2.02 \times 10^{16}$ | $1.56 \times 10^{16}$ | $3.06 \times 10^{15}$ | $2.43 \times 10^{14}$ |
|    | 35 | $2.53 \times 10^{16}$ | $1.95 \times 10^{16}$ | $3.49 \times 10^{15}$ | $3.10 \times 10^{14}$ |
|    | 40 | $3.07 \times 10^{16}$ | $2.31 \times 10^{16}$ | $5.02 \times 10^{15}$ | $5.89 \times 10^{14}$ |
| 48 | 5  | $1.29 \times 10^{16}$ | $5.23 \times 10^{15}$ | $7.33 \times 10^{15}$ | 0 |
|    | 10 | $1.64 \times 10^{16}$ | $1.14 \times 10^{16}$ | $3.99 \times 10^{15}$ | $3.33 \times 10^{13}$ |
|    | 15 | $1.96 \times 10^{16}$ | $1.46 \times 10^{16}$ | $3.45 \times 10^{15}$ | $1.47 \times 10^{14}$ |
|    | 20 | $2.31 \times 10^{16}$ | $1.83 \times 10^{16}$ | $3.64 \times 10^{15}$ | $3.32 \times 10^{14}$ |
|    | 25 | $3.24 \times 10^{16}$ | $2.42 \times 10^{16}$ | $5.35 \times 10^{15}$ | $2.94 \times 10^{14}$ |
|    | 30 | $3.63 \times 10^{16}$ | $2.59 \times 10^{16}$ | $7.73 \times 10^{15}$ | $3.11 \times 10^{14}$ |
|    | 35 | $4.84 \times 10^{16}$ | $3.46 \times 10^{16}$ | $1.04 \times 10^{16}$ | $5.23 \times 10^{14}$ |
|    | 40 | $6.17 \times 10^{16}$ | $4.36 \times 10^{16}$ | $1.32 \times 10^{16}$ | $5.87 \times 10^{14}$ |
| 96 | 5  | $2.44 \times 10^{16}$ | $1.00 \times 10^{16}$ | $1.37 \times 10^{16}$ | 0 |
|    | 10 | $3.28 \times 10^{16}$ | $2.31 \times 10^{16}$ | $7.62 \times 10^{15}$ | $1.53 \times 10^{14}$ |
|    | 15 | $4.68 \times 10^{16}$ | $3.41 \times 10^{16}$ | $1.00 \times 10^{16}$ | $2.55 \times 10^{14}$ |
|    | 20 | $5.74 \times 10^{16}$ | $4.26 \times 10^{16}$ | $1.27 \times 10^{16}$ | $6.67 \times 10^{14}$ |
|    | 25 | $8.13 \times 10^{16}$ | $5.72 \times 10^{16}$ | $1.81 \times 10^{16}$ | $1.12 \times 10^{15}$ |
|    | 30 | $9.24 \times 10^{16}$ | $6.51 \times 10^{16}$ | $1.96 \times 10^{16}$ | $1.47 \times 10^{15}$ |
|    | 35 | $1.13 \times 10^{17}$ | $8.27 \times 10^{16}$ | $2.46 \times 10^{16}$ | $1.87 \times 10^{15}$ |
|    | 40 | $1.28 \times 10^{17}$ | $9.33 \times 10^{16}$ | $2.46 \times 10^{16}$ | $2.60 \times 10^{15}$ |

## S-3. Supplementary figures

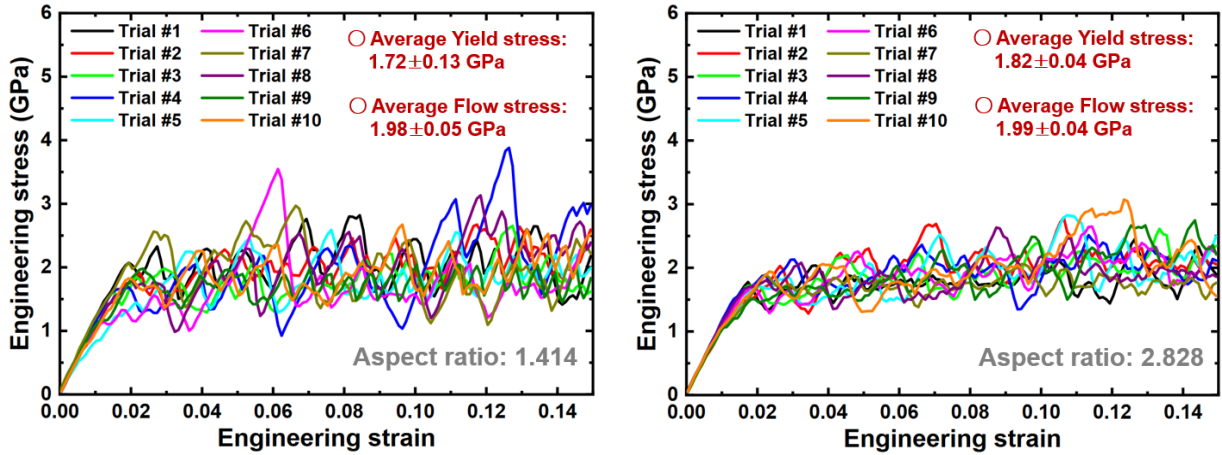

Fig. S1  The aspect ratio (= height/width) dependence of the stress-strain responses of the 21 nm sized [123] TCP at 300 K. The pillar was prepared by the vacancy annealing method with an initial vacancy concentration of 10 % and heat treated for 750 ps at 1100 K before the loading.

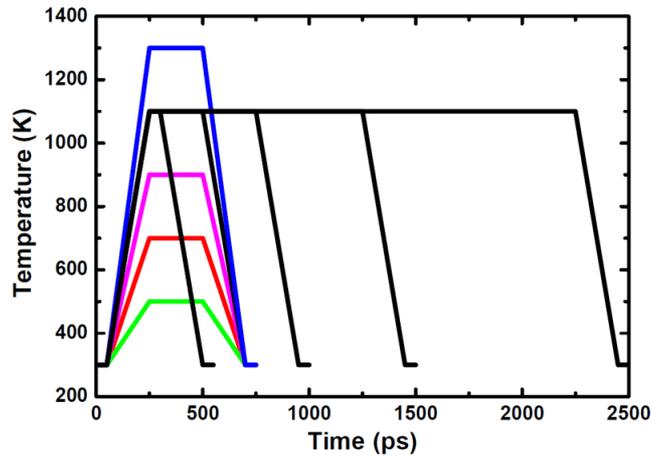

Fig. S2  Illustration of the heat treatment conditions used to introduce a preexisting (prior to the deformation), randomly distributed network of dislocations inside of the nanopillars.

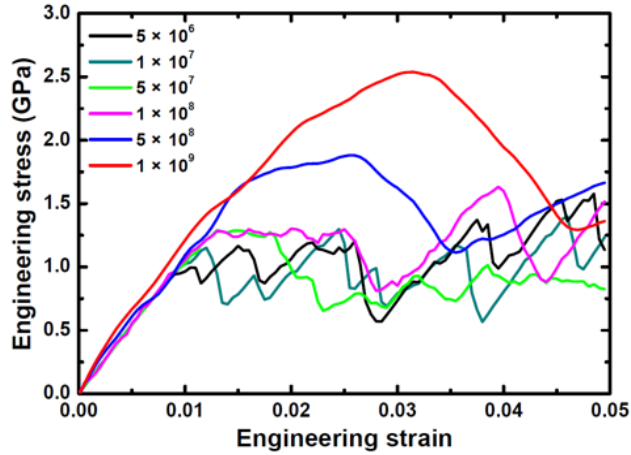

Fig. S3  Strain rate dependence of the stress-strain response of the 28 nm sized [123] SCP at 300 K. The pillar was prepared by the vacancy annealing method with an initial vacancy concentration of 10 % and heat treated for 750 ps at 1100 K before the loading.

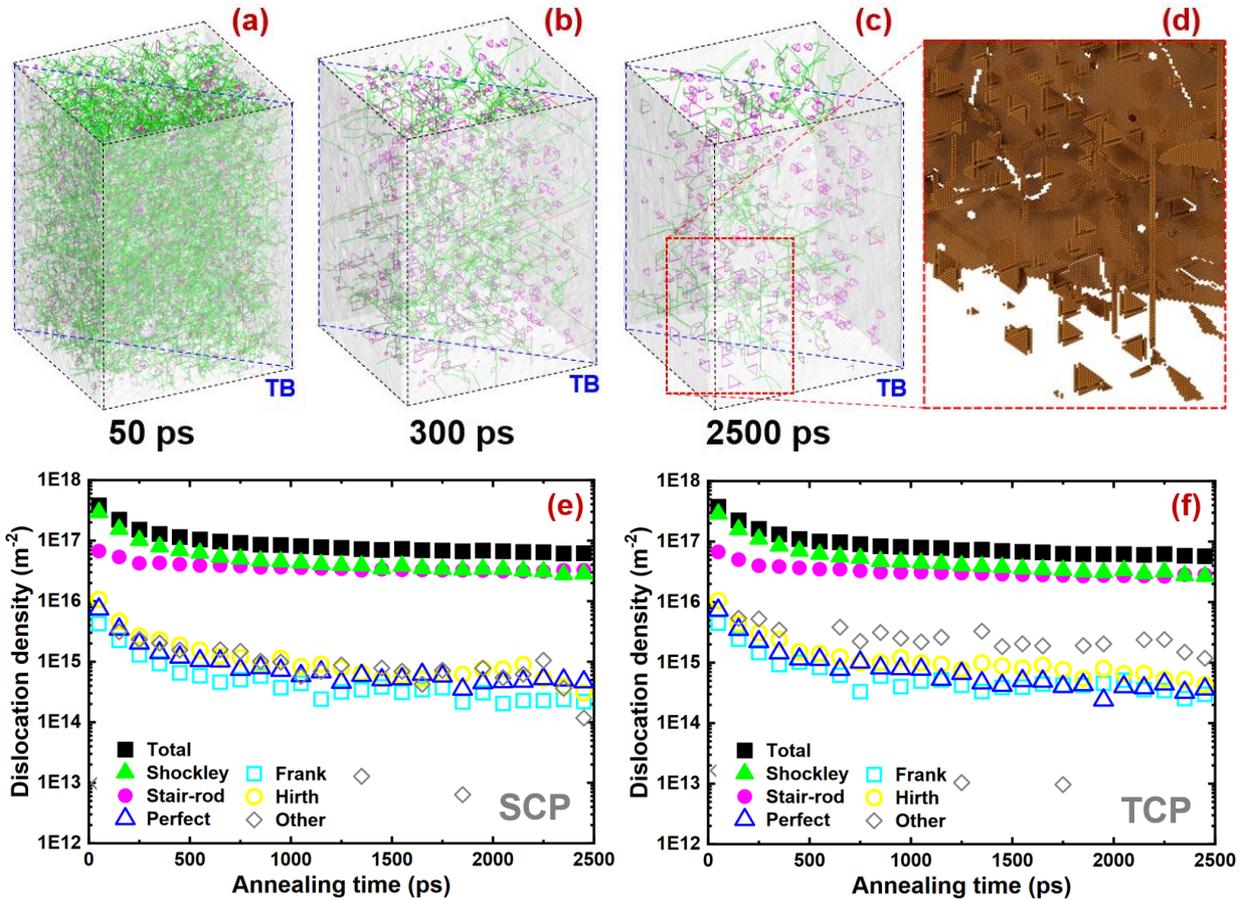

Fig. S4  Evolution of the dislocation structure during heat treatment at 1100 K in the 42 nm sized [101] pillars with an initial vacancy concentration of 10 %. Snapshots of the TCP at (a) 50, (b) 300, and (c) 2500 ps are shown. The color of the lines is scaled according to the DXA, with green lines corresponding to Shockley partials, magenta lines to stair-rod dislocations, and blue lines to perfect dislocations. (d) Zoom into (c) highlighting that the stair rod dislocations form stacking fault tetrahedra made of hexagonal-close packed (hcp) atoms colored according to the Ackland–Jones analysis [18]. Density of the various dislocation components as a function of the annealing time for an (e) SCP and (f) TCP.

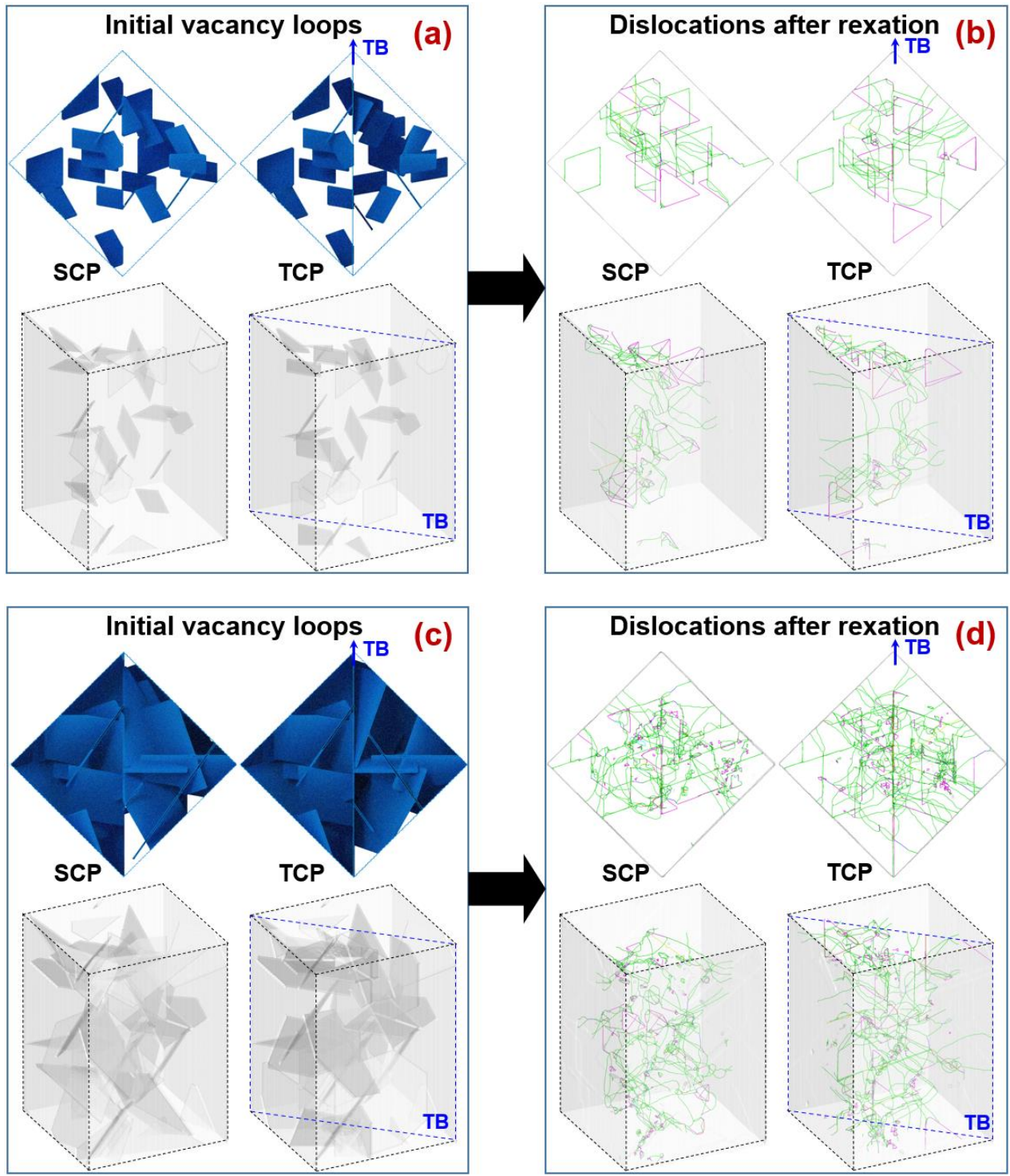

Fig. S5  Evolution of the dislocation structure in the 42 nm sized [101] SCPs and TCPs with initially 24 vacancy loops. The widths of the dislocation loops are (a, b) 15 × *a* (0.3615 nm) and (c, d) 40 × *a* (0.3615 nm). Atomic configurations (a, c) before and (b, d) after relaxation at 300 K for 100 ps are presented. In the dislocation structures after the relaxation, the color of the lines is scaled according to the DXA, with green lines corresponding to Shockley partials, magenta lines to stair-rod dislocations, and blue lines to perfect dislocations.

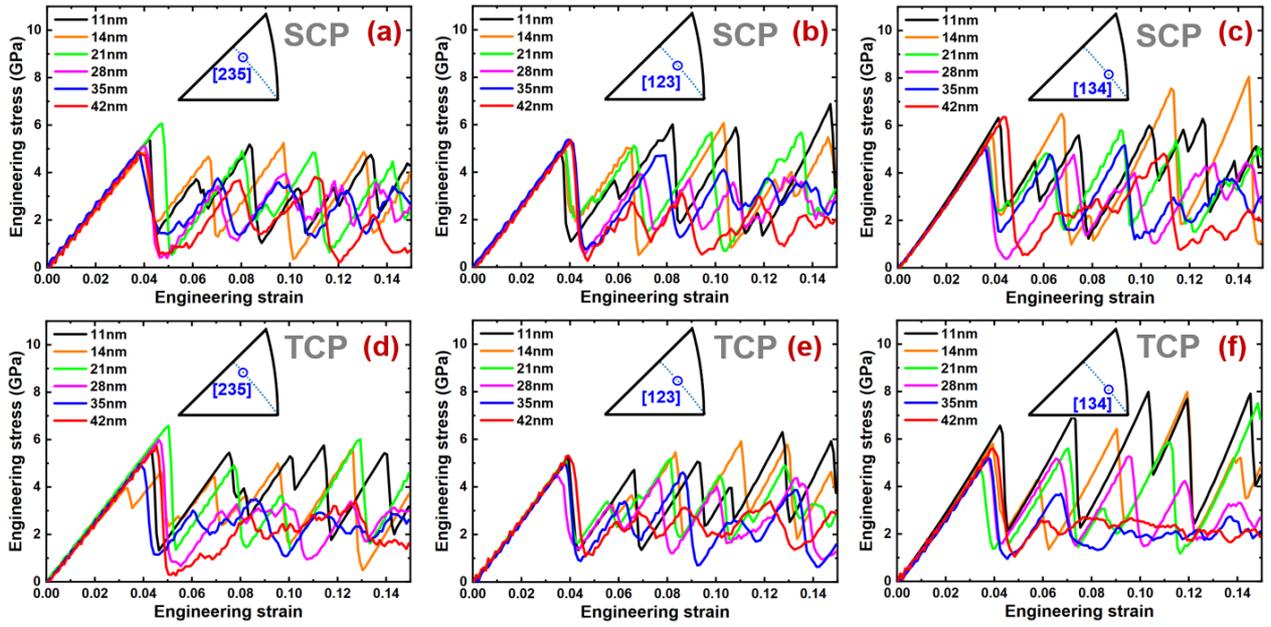

Fig. S6  Stress-strain response of initially dislocation-free (a, b, c) SCPs and (d, e, f) TCPs at 300 K. Results for various pillar sizes (11, 14, 21, 28, 35, and 42 nm) and different loading orientations of (a, d) [235], (b, e) [123], and (c, f) [134] are presented.

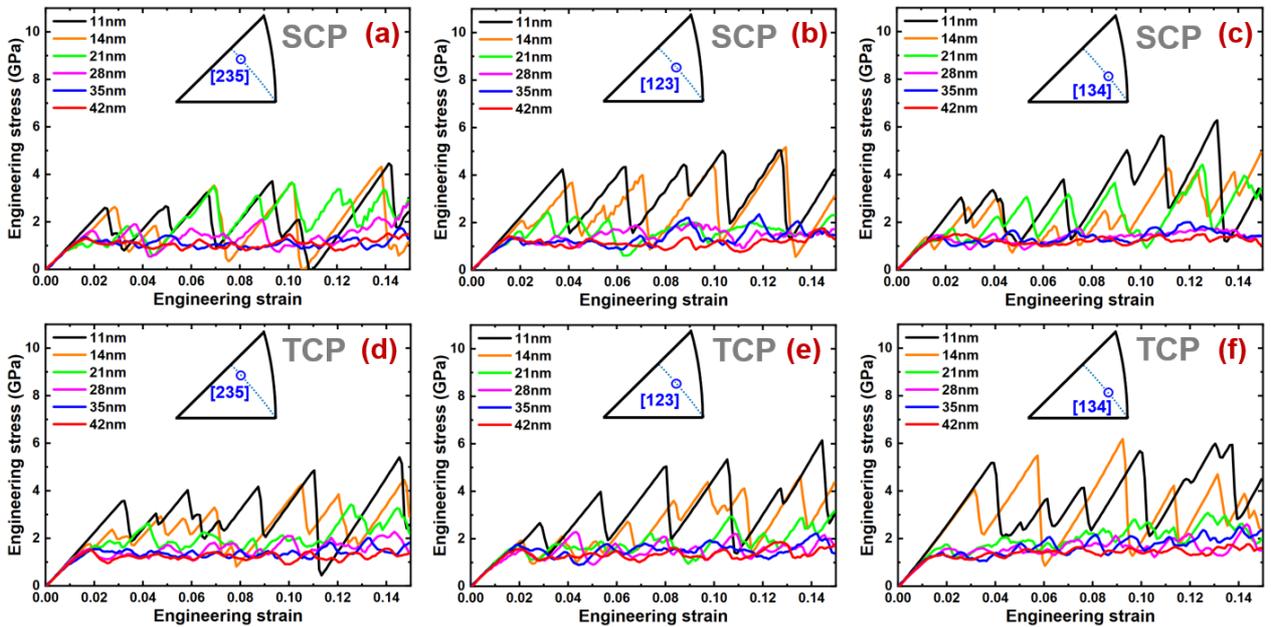

Fig. S7  Stress-strain response at 300 K of (a, b, c) SCPs and (d, e, f) TCPs prepared by a heat treatment of 750 ps at 1100 K with an initial vacancy concentration of 10 %. Results for various pillar sizes (11, 14, 21, 28, 35, and 42 nm) and different loading orientations of (a, d) [235], (b, e) [123], and (c, f) [134] are presented.

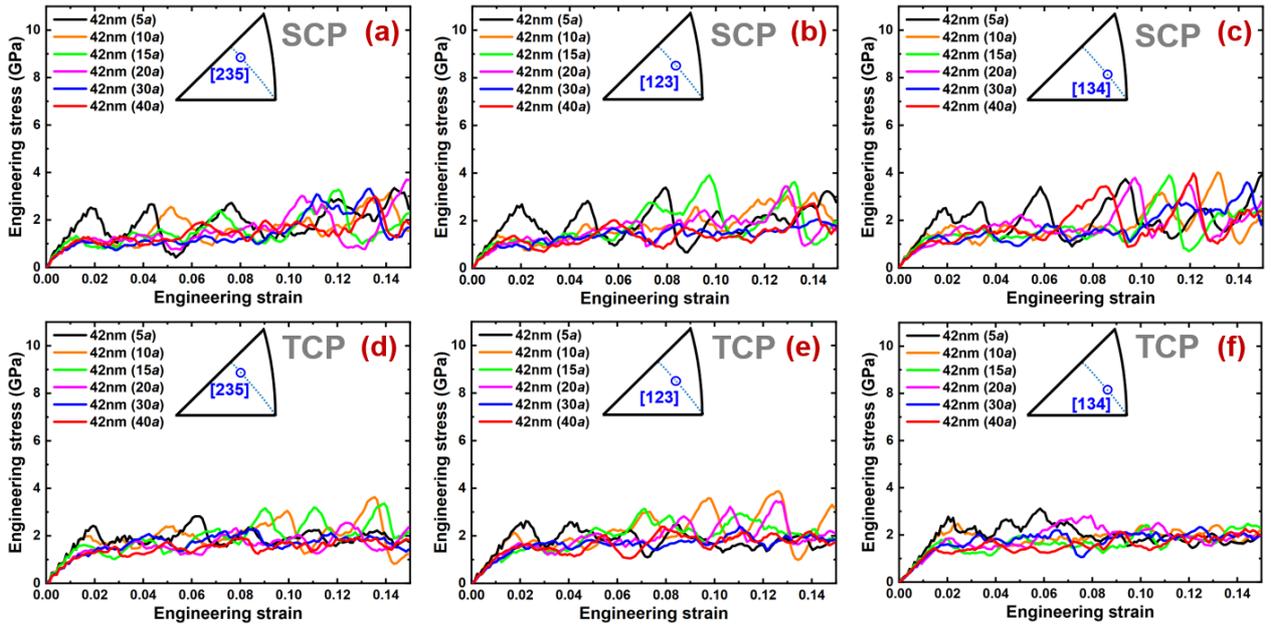

Fig. S8 Stress-strain response at 300 K of 42 nm sized (a, b, c) SCPs and (d, e, f) TCPs prepared by a relaxation time of 100 ps at 300 K with initially 24 vacancy loops with different widths of (5, 10, 15, 20, 30, and 40) × $a$ (0.3615 nm). Results for different loading orientations of (a, d) [235], (b, e) [123], and (c, f) [134] are presented.

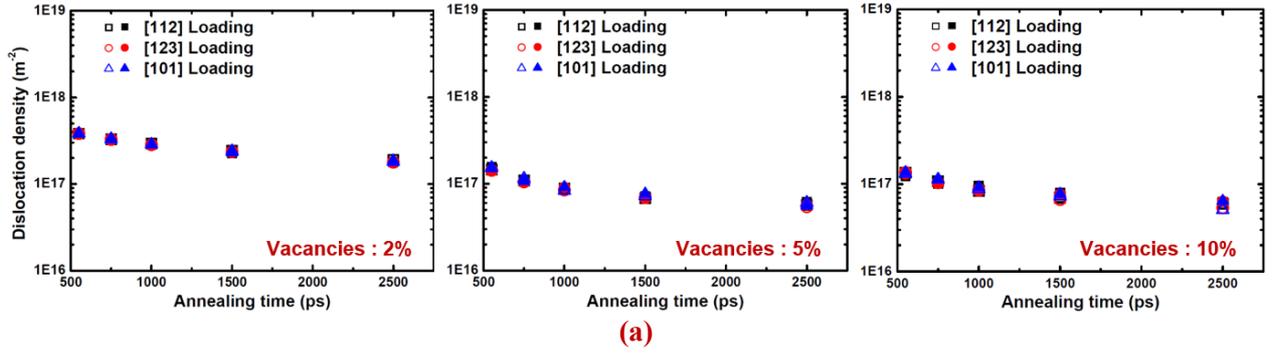
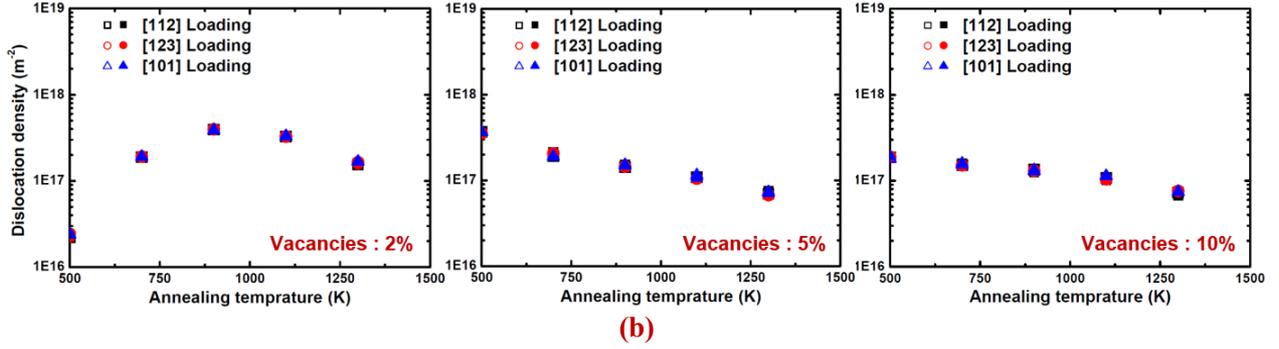

Fig. S9  Total dislocation density as a function of (a) the annealing time (heat treated at 1100 K) and (b) annealing temperature (heat treated for 750 ps). Vacancy concentrations of 2, 5, and 10 % (from left to right) were introduced into the 42 nm sized pillars before the heat treatment. Open and filled symbols represent values for SCPs and TCPs, respectively.

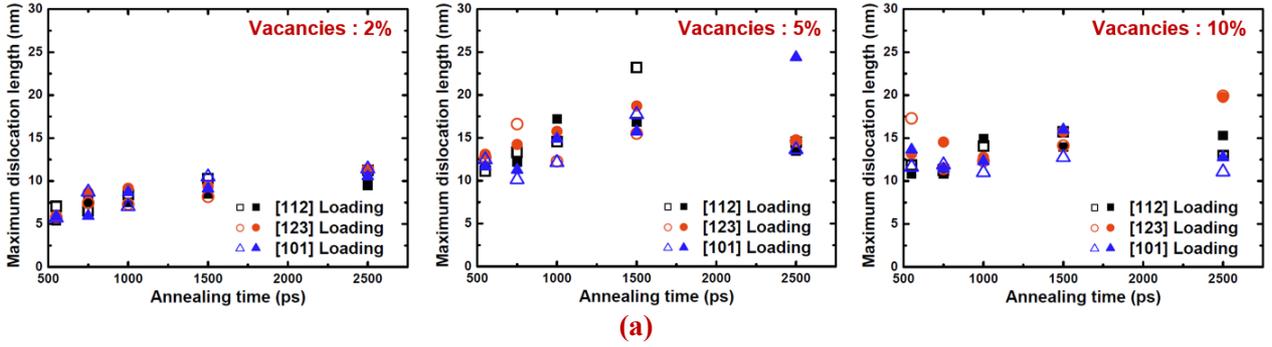
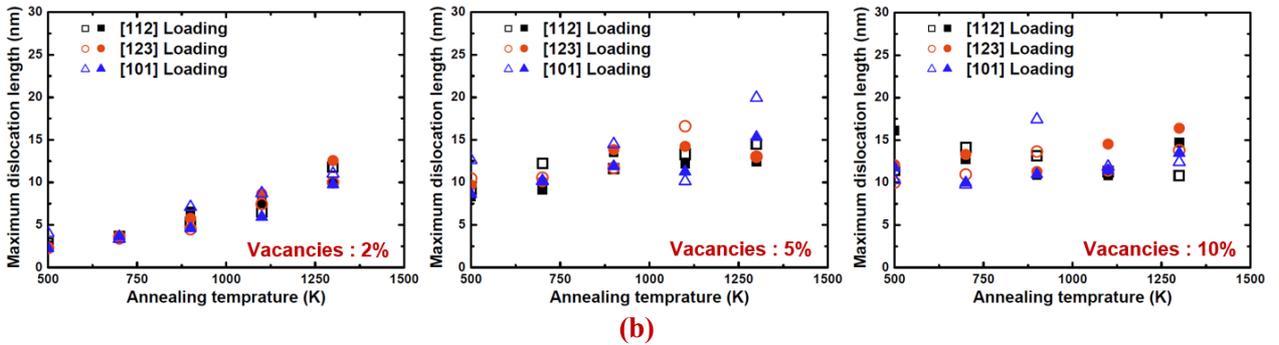

Fig. S10  Maximum dislocation length as a function of (a) the annealing time (heat treated at 1100 K) and (b) annealing temperature (heat treated for 750 ps). Vacancy concentrations of 2, 5, and 10 % (from left to right) were introduced into the 42 nm sized pillars before the heat treatment. Open and filled symbols represent values for SCPs and TCPs, respectively.

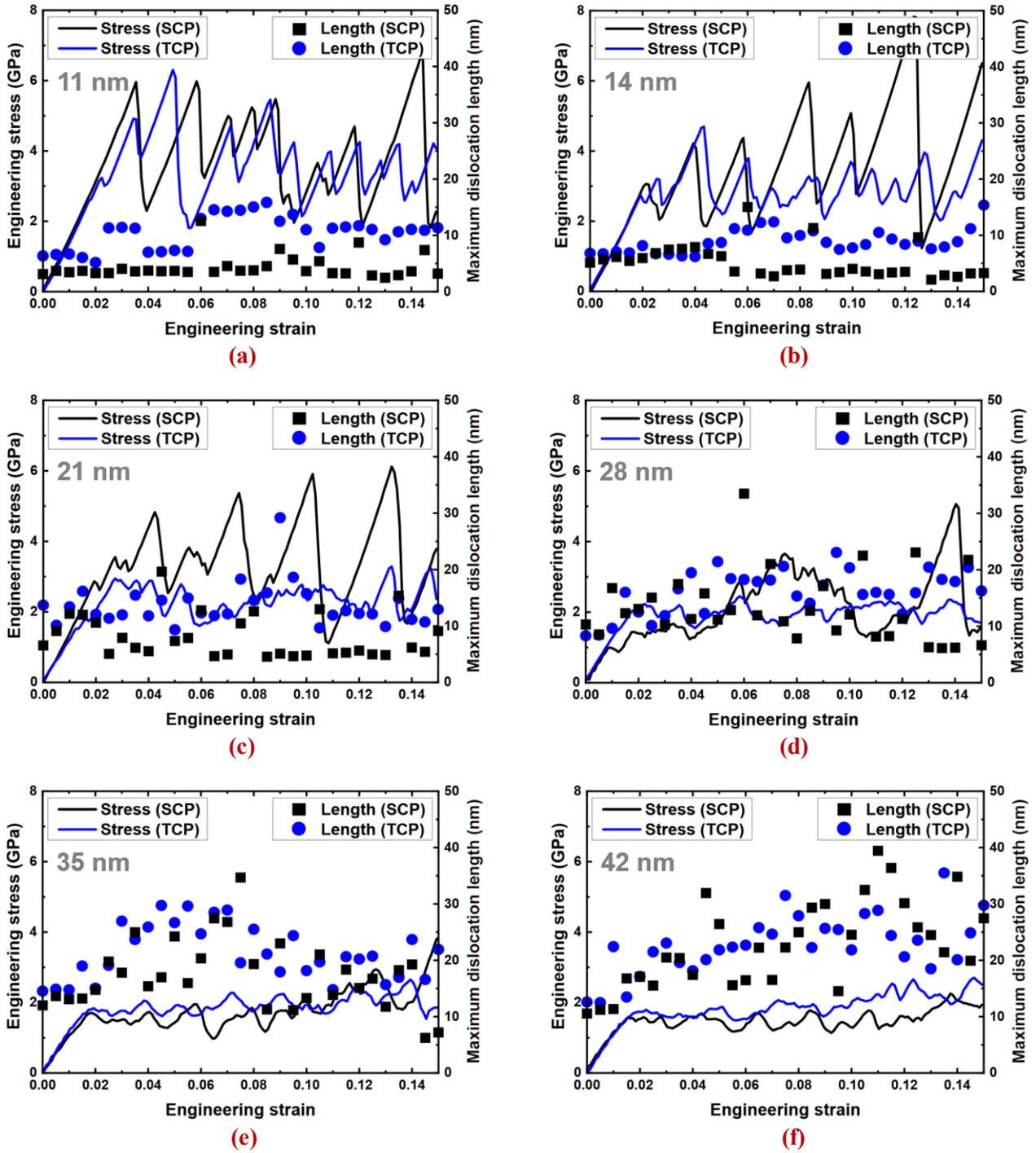

Fig. S11 Correlation between the flow stress and the maximum dislocation length for the (a) 11, (b) 14, (c) 21, (d) 28, (e) 35, and (f) 42 nm sized [101] SCPs and (b) TCPs during compressive loading at 300 K based on the EAM potential by Sheng *et al.* [7]. Each pillar was prepared by the vacancy annealing method with an initial vacancy concentration of 10 % and heat treated for 750 ps at 1100 K before loading.